\documentclass[aps,pre,reprint,superscriptaddress]{revtex4-2}

\usepackage{amsmath}
\usepackage{graphicx}
\usepackage{bm}
\usepackage{siunitx}
\usepackage{placeins}

\begin{document}

\title{The Effective Gravitational Field of the Ball in Association Football}

\author{Oliver~B.~Wright}
\email{olly@eng.hokudai.ac.jp}
\affiliation{Plum Science Co., Ltd., Haruno Bldg. 3F, 1-323 Minami 1 Nishi 16, Chuo-ku, Sapporo 060-0061, Japan}
\affiliation{Hokkaido University, Sapporo 060-0808, Japan}
\affiliation{Graduate School of Engineering, The University of Osaka, 2-1 Yamadaoka, Suita 565-0871, Japan}

\date{\today}

\begin{abstract}
In association football, collective player motion is organized around the
ball. We ask whether this many-agent motion can be described by an
effective field analogous to gravitational attraction, emphasizing that
the measured quantity is a radial drift velocity, as in overdamped
dynamics, rather than a Newtonian acceleration. Using public tracking data
from ten matches, we characterize this radial drift velocity through the
distance-dependent mean $D(r)$ and the ball-position-dependent field
$K(x,y)$. After excluding restarts and other constrained phases of play,
both goalkeepers, the player nearest to the ball, and small player--ball separations, a persistent inward drift velocity remains, with a
global mean of about $0.9~\mathrm{m\,s^{-1}}$. The distance dependence
is not described by an inverse power of $r$: $D(r)$ decreases at short
range and then forms a broad plateau that persists to the largest measured
separations. The mean $K(x,y)$ is positive over most of the pitch, with a
central depression about $15\%$ below the non-goalmouth average and
stronger suppression near the goalmouths, where the mean drift velocity
becomes slightly negative, corresponding to weak effective repulsion.
The role-resolved $D(r)$ shows a pronounced defender minimum near $r\simeq12~\mathrm{m}$ and a pronounced attacker maximum near $r\simeq60~\mathrm{m}$, whereas only defenders produce effective repulsion near the goals in $K(x,y)$. These results show that football tracking data can reveal
simple effective laws of active many-body motion while retaining clear
signatures of player role and pitch geometry.
\end{abstract}

\maketitle

\section{Introduction}

People often say that when amateurs play football, the players look like bees around a honeypot. The phrase is meant as criticism, but it also contains a physical question. The ball is a moving focus of attention, and the players continually reorganize their positions around it. At low skill levels the attraction may look like a collapse onto the ball. At higher levels the player--ball distances are larger, the structure is more organized, and tactical constraints are stronger, but the ball still acts as the dominant moving object in the game. The question is therefore not whether players follow the ball, but how this motion can be quantified.

The gravitational analogy is useful because it gives a simple language for the problem. In Newtonian gravity, a mass generates a field, and other bodies accelerate in response to it. In football, however, the analogy cannot be literal. Players are active, decision-making, speed-limited agents, not passive particles, and the relevant response is not a sustained acceleration toward the ball. The more natural quantity to measure is an overdamped drift velocity: after reaction and decision processes have occurred, how much of a player's velocity is directed toward the instantaneous ball position? Similar effective-force or effective-field descriptions have been useful in other many-agent systems, including pedestrian dynamics and active matter \cite{Helbing1995,Vicsek1995,Toner1995}. Experiments on animal groups, including the murmurations of starling flocks, have also shown that collective response must be quantified without reducing each agent to a simple passive body \cite{Ballerini2008,Cavagna2010}.

Football is already a major application area for spatiotemporal tracking data. Such data have supported large-scale match analysis and surveys of tactical and positional analysis in team sports \cite{Bialkowski2014,Gudmundsson2017,Low2020,ReinMemmert2016,Goes2021}. More specifically, tracking-based football studies have used player and ball positions to estimate which player or team can reach each part of the pitch first, how open or dangerous different locations are, and how valuable a pass into a given region may be \cite{TakiHasegawa2000,FujimuraSugihara2005,FernandezBornn2018,Spearman2018,Martens2021,Link2016,Power2017,Narizuka2021}. Novillo et al. \cite{Novillo2024} analyzed the velocity modulus, i.e., the total speed, and the angle of the player velocity relative to the ball direction as functions of player--ball distance, player role, and attacking or defensive phase of play. In parallel, physics- and dynamics-oriented studies have examined statistical regularities, the motion and spacing of team formations, player-motion correlations, ball-possession dynamics, marking dynamics, and stochastic interaction models \cite{Mendes2007,Kijima2014,NarizukaYamazaki2017,Frencken2011,Moura2013,Welch2021,Chacoma2020,ChacomaMarking2022,Chacoma2021}. Despite this extensive work, to our knowledge tracking data have not previously been used to extract the signed ball-directed drift velocity of the players either as an effective distance-dependent radial law $D(r)$ or as a ball-position map of the same inward radial drift velocity, $K(x,y)$.

We address this gap with a direct kinematic approach. Using publicly
available match tracking data \cite{MetricaData,Bundesliga}, we
apply the method to ten games. The retained sample is restricted to
periods of open play, with the goalkeepers and the player nearest to the
ball excluded. At each retained time, we calculate the component of each
included player's velocity toward the instantaneous ball position. For
the eight games with registered player roles, we also resolve the
analysis into attackers, midfielders, defenders and goalkeepers.

Averaging this inward velocity component at fixed player--ball distance gives the
radial drift velocity $D(r)$, whereas averaging it at fixed ball position
gives the drift velocity field $K(x,y)$. Together, these quantities can be used to test the
gravity analogy without assuming a particular functional form. Within this
analogy, positive mean drift velocity corresponds to effective attraction toward the
ball, whereas negative mean drift velocity corresponds to effective repulsion.

\section{Radial drift velocity and effective attraction}

The gravitational analogy suggests looking for a field generated by the ball. In football, the quantity of interest is not a Newtonian acceleration. Players are self-propelled agents, and their motion is limited by reaction time, running speed, anticipation, and tactics. The simplest useful quantity is therefore the component of a player's velocity directed toward the instantaneous ball position. 

\begin{figure}[t]
\centering
\includegraphics[width=0.95\linewidth]{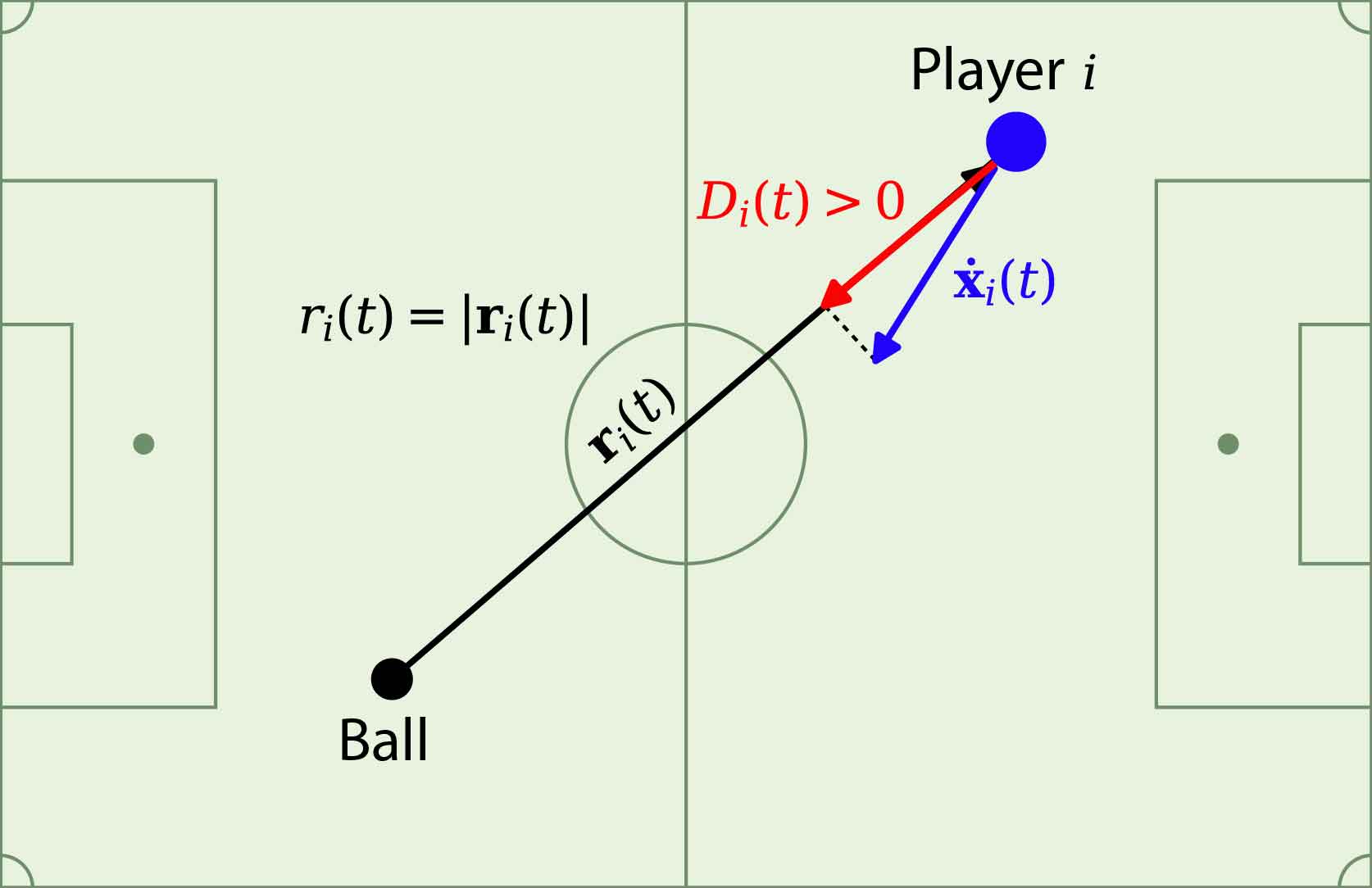}
\caption{Definition of the radial drift velocity. The vector $\mathbf{r}_i(t)$ points from the ball to player $i$. The player velocity $\dot{\mathbf{x}}_i(t)$ is projected onto this radial direction, with the sign chosen so that positive $D_i(t)$ corresponds to motion toward the ball.}
\label{fig1}
\end{figure}

Let $\mathbf{b}(t)$ be the ball position and $\mathbf{x}_i(t)$ the position of player $i$ at time $t$. The vector from the ball to the player is
\begin{equation}
\mathbf{r}_i(t)=\mathbf{x}_i(t)-\mathbf{b}(t),
\qquad
r_i(t)=|\mathbf{r}_i(t)|.
\label{eq:player_ball_vector}
\end{equation}
Here $r_i(t)$ is the two-dimensional distance on the pitch between player $i$ and the supplied ball coordinate. Thus, when the ball is airborne, the distance is computed using its projection on the pitch plane rather than its three-dimensional position. The player velocity is estimated from the tracking data by the central difference
\begin{equation}
\dot{\mathbf{x}}_i(t)\simeq
\frac{\mathbf{x}_i(t+\Delta t)-\mathbf{x}_i(t-\Delta t)}
{2\Delta t}.
\label{eq:velocity}
\end{equation}
The radial drift velocity of player $i$ toward the ball is then defined as
\begin{equation}
D_i(t)=
-\dot{\mathbf{x}}_i(t)\cdot
\frac{\mathbf{r}_i(t)}{r_i(t)}.
\label{eq:inward_drift}
\end{equation}
This definition and sign convention are illustrated in Fig.~\ref{fig1}. The minus sign makes the convention intuitive. If $D_i(t)>0$, the player is moving toward the ball. If $D_i(t)<0$, the player is moving away from it. The quantity $D_i(t)$ has units of velocity, so it should be interpreted as an overdamped drift velocity rather than as a force or acceleration. \footnote{An alternative scalar quantity, $-\dot r_i(t)$, estimated with the same
central difference, measures the rate of change of the player--ball
separation and therefore includes the motion of the ball. Its equal-weight
10-game mean lies between about $-0.14$ and $+0.01~\mathrm{m\,s^{-1}}$
over $2\le r<65~\mathrm{m}$, consistent with little systematic change in
player--ball separation on average. We retain Eq.~\eqref{eq:inward_drift} because it measures
the player's directed motion toward the instantaneous ball position.}

Two averages of Eq.~\eqref{eq:inward_drift} are used below. The first characterizes the player--ball distance-dependent ``gravitational'' law. Let $\text{bin}(r)$ denote a one-dimensional interval of player--ball distance centered at $r$. If $N_r$ is the number of recorded player--time observations with $r_i(t)\in\text{bin}(r)$, then
\begin{equation}
D(r)=
\frac{1}{N_r}
\sum_{\substack{(i,t):\ r_i(t)\in\text{bin}(r)}}
D_i(t).
\label{eq:radial_drift}
\end{equation}
The function $D(r)$ is the mean inward drift velocity as a function of player--ball distance $r$, averaged over player index $i$ and time $t$ within that distance interval. It is the nearest analog, in the present overdamped setting, of a radial attraction law: as in motion through a viscous medium under a constant force, the relevant physical quantity is a drift velocity rather than a continuing acceleration.

The second average represents how the attraction strength varies with ball position on the pitch. Unlike the ideal Newtonian problem of a point mass in homogeneous space, the football pitch is not a uniform universe: the average player motion around a ball near a touchline, corner, or goalmouth need not be the same as around a ball in midfield. Dividing the pitch into rectangular cells, let $\text{cell}(x,y)$ denote the cell centered at $(x,y)$, where $x$ is measured along the goal-to-goal direction and $y$ across the width of the pitch. If $N_{xy}$ is the number of recorded player--time observations for which the ball lies in this cell, then
\begin{equation}
K(x,y)=
\frac{1}{N_{xy}}
\sum_{\substack{(i,t):\ \mathbf{b}(t)\in\text{cell}(x,y)}}
D_i(t).
\label{eq:K_field}
\end{equation}
The field $K(x,y)$, which we term the ball-position drift velocity field, is the mean inward drift velocity, expressed as a function of ball position. In this average the player--ball distance $r$ is not fixed; the average is taken over the player positions that occur whenever the ball is in the specified cell, and therefore over the corresponding times $t$ as well. Its role is analogous to a local strength parameter of attraction, with
$K(x,y)<0$ corresponding to local effective repulsion. If the radial law took the form $D(r)\propto G/r^2$ or $D(r)\propto G/r$, for example, the natural quantity to map would be the fitted coefficient $G$ across the pitch. Determining whether this football ``gravitation'' strength varies across the pitch is a core objective of this paper.

\section{Tracking data and included play}
\label{sec:data}
The analysis is based on ten games corresponding to public Metrica Sports tracking data for three anonymized matches \cite{MetricaData} and Bundesliga tracking data for seven identified matches \cite{Bundesliga}. In all cases, player and ball coordinates were converted to meters on a $105\times68~\mathrm{m}$ pitch, and the tracking interval was $\Delta t=0.04~\mathrm{s}$ for both data sets. The Metrica coordinates correspond to increments of about 1~$\mathrm{mm}$ on the pitch, whereas the Bundesliga coordinates are given in 1~$\mathrm{cm}$ increments. These values
describe only the numerical precision of the stored coordinates and
should not be interpreted as the physical tracking accuracy or as an
exact measurement of a player's center of mass.

The analysis started from the full-game tracking files. The retained sample is smaller because the aim is to characterize open-play motion toward the ball, not motion during restarts or other constrained situations. The raw data were processed with a Python analysis pipeline, which identified unusable time intervals and discarded the corresponding tracking frames before any drift velocities were calculated. Frames were excluded when the ball position was missing or outside the pitch, when neighboring tracking frames required for the evaluation of Eq.~\eqref{eq:velocity} were unavailable, or when play was associated with corners, throw-ins, goal kicks, kickoffs, free kicks, fouls, or similar stoppages. The same filtering logic was applied to all ten games: frames during and around constrained phases of play were removed before calculating $D_i(t)$. The detailed numerical rule and its application to the two
data sets are given in Appendix~\ref{app:filtering}.

At each retained time, the analysis pipeline identified the player closest to the ball and omitted that player before calculating the averages. This rule uses only the computed player--ball distances; it does not require defining which team has possession or which player is controlling the ball. The closest player is often directly involved with the ball, for example by dribbling, receiving, or challenging for it. Very small player--ball separations, $r<2~\mathrm{m}$, were also excluded. This removes the immediate ball-interaction region, where contact-scale events---touches, tackles, shielding, and rebounds---dominate, rather than the motion of the surrounding players. The public tracking files provide a single two-dimensional ball coordinate in each frame. Such data are well-suited for locating the ball on the pitch and constructing the tactical-scale drift velocity field considered here, but not for resolving ball--foot contact or close-control microdynamics. The effect of retaining the closest player is examined in Appendix~\ref{app:nearest_player_check}; it changes only the shortest-distance part of $D(r)$ slightly and has little effect on the overall radial law or the field $K(x,y)$.

\begin{figure}[t]
\centering
\includegraphics[width=0.95\linewidth]{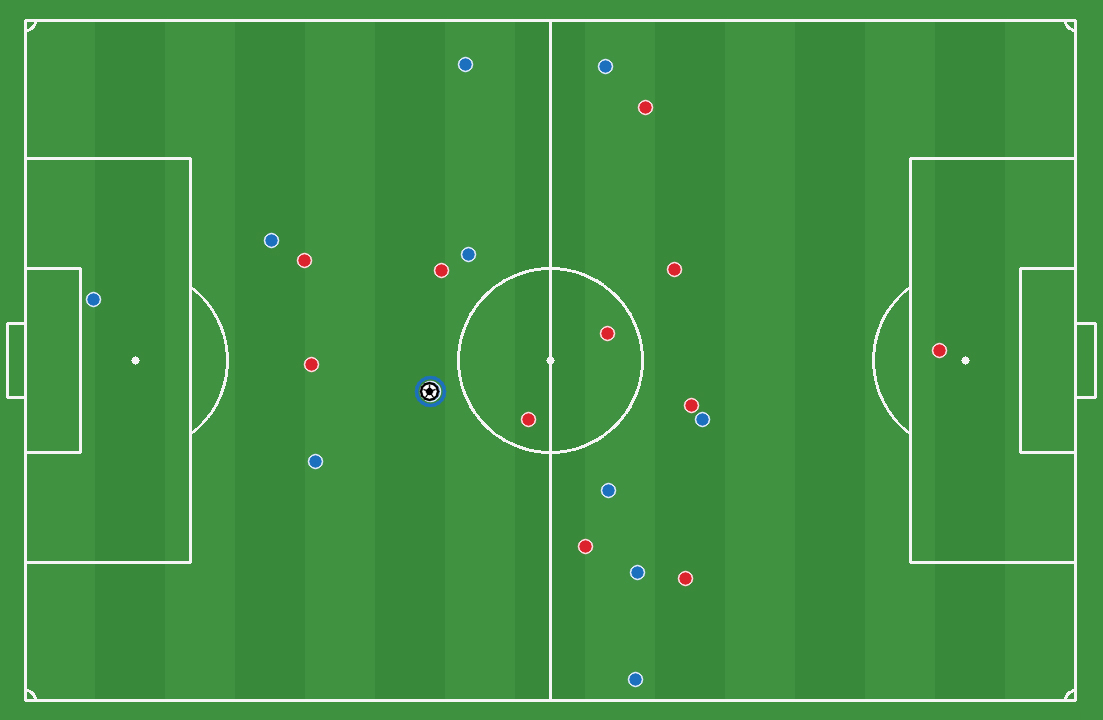}
\caption{Typical snapshot of the ball and player positions in Game~1. The opposing teams are represented by red (Away) and blue (Home). The enlarged blue marker identifies the Home player closest to the ball. The ball is shown by the black-and-white marker and lies very close to that player in this frame, with a separation of $r=0.13~\mathrm{m}$. The frame corresponds to $t=2723.44~\mathrm{s}$ ($\simeq45.4~\mathrm{min}$) in period~1. A $20~\mathrm{s}$ movie clip including this frame is given in Supplementary Video~1.}
\label{fig:mov}
\end{figure}

Figure~\ref{fig:mov} shows a typical still from Game~1. The blue player closest to the ball is shown by an enlarged marker, and the two teams are distinguished by blue and red. A $20~\mathrm{s}$ video of a typical midfield situation from Game~1 is given in Supplementary Video~1.

\begin{table*}[t]
\caption{Size of the 10-game data set after the exclusions described in the text. The half-time score is given in the same team order as the final score. The last column gives the number of individual values of $D_i(t)$ used in the averages. Games~1--3 use Metrica Sports data. For Games~1 and 2, the two sides are labeled Home and Away, and for Game~3 they are labeled Team~A and Team~B. Games~4--10 use Bundesliga data.}
\label{tab:data_summary}
\begin{ruledtabular}
\begin{tabular}{llccc}
Game & Match and final score & Half-time score & Included time & Values of $D_i(t)$ \\
\hline
Game 1 & Home 3--1 Away
& 1--0 & $39.95~\mathrm{min}$ & $1{,}132{,}657$ \\
Game 2 & Home 3--2 Away
& 1--1 & $37.75~\mathrm{min}$ & $1{,}068{,}920$ \\
Game 3 & Team A 0--2 Team B
& 0--1 & $35.55~\mathrm{min}$ & $1{,}008{,}516$ \\
Game 4 & 1. FC K\"oln 1--2 FC Bayern M\"unchen
& 0--1 & $38.31~\mathrm{min}$ & $1{,}086{,}006$ \\
Game 5 & Fortuna D\"usseldorf 4--0 SSV Jahn Regensburg
& 0--0 & $30.31~\mathrm{min}$ & $859{,}441$ \\
Game 6 & VfL Bochum 1848 3--0 Bayer 04 Leverkusen
& 2--0 & $26.72~\mathrm{min}$ & $718{,}744$ \\
Game 7 & Fortuna D\"usseldorf 3--1 F.C. Hansa Rostock
& 2--0 & $32.02~\mathrm{min}$ & $908{,}055$ \\
Game 8 & Fortuna D\"usseldorf 0--1 1. FC N\"urnberg
& 0--0 & $36.84~\mathrm{min}$ & $1{,}045{,}026$ \\
Game 9 & Fortuna D\"usseldorf 1--0 FC St. Pauli
& 1--0 & $40.61~\mathrm{min}$ & $1{,}138{,}256$ \\
Game 10 & Fortuna D\"usseldorf 1--2 1. FC Kaiserslautern
& 1--0 & $41.56~\mathrm{min}$ & $1{,}180{,}200$ \\
\hline
Total & -- & -- & $359.62~\mathrm{min}$ & $10{,}145{,}821$ \\
\end{tabular}
\end{ruledtabular}
\end{table*}

In all ten games, player velocities were calculated from the central
difference in Eq.~\eqref{eq:velocity}, using the positions at
$t+\Delta t$ and $t-\Delta t$ with $\Delta t=0.04~\mathrm{s}$. The
two positions used for each velocity are separated by
$0.08~\mathrm{s}$. Although differentiation can in principle amplify
short-time fluctuations in the measured positions, we verified that
the resulting velocity statistics are insensitive to such fluctuations
at the level relevant here: the mean player speeds changed by less than
$0.2\%$ in representative tests using smoothed trajectories. Thus the
central-difference estimates are sufficiently stable for the present
analysis, in which large numbers of individual values are subsequently
averaged. We then calculated each retained value of $D_i(t)$. To obtain $D(r)$,
these values were averaged in successive $5~\mathrm{m}$ distance
intervals beginning at $r=0$. Observations with $r<2~\mathrm{m}$ were
excluded, so the first interval, $0$--$5~\mathrm{m}$, contains only
observations with $2\le r<5~\mathrm{m}$. For a fixed ball position, each distance interval
corresponds to an annular region centered on the ball; near the pitch
boundaries, and at sufficiently large $r$, this region is truncated by
the finite rectangular pitch. The $5~\mathrm{m}$ interval width is not
the spatial resolution of the tracking data, but only the width over
which values of $D_i(t)$ are averaged. A much finer interval, such as
$1~\mathrm{m}$, would give noisier averages and is not needed for the
question addressed here, namely the large-scale form of $D(r)$.

To obtain $K(x,y)$, the same values of $D_i(t)$ were averaged separately for each rectangular pitch cell containing the ball. The full-pitch grid contains $24\times16$ cells; with the pitch dimensions stated above, each cell has dimensions $4.375\times4.25~\mathrm{m}$. For the main $K(x,y)$ maps, equivalent cells were symmetry-averaged to improve statistics and display the common pitch-scale pattern. The main results below use the same definitions, exclusion rule, time step, distance intervals, and pitch grid for all ten games. Further numerical details are given in Appendix~\ref{app:filtering}; the maps before symmetry averaging are included in Appendix~\ref{app:unfolded_team}.

\section{Distance dependence of the attraction}
\label{sec:radial_result}
We first consider the distance-dependent radial drift velocity $D(r)$. This is the most direct test of the gravitational analogy, because in Newtonian gravity the field of a point source is identified by how its strength varies with distance from that source. In the football analogy, the source is the ball, and $r$ is the player--ball separation.

\begin{figure}[t]
\centering
\includegraphics[width=0.95\linewidth]{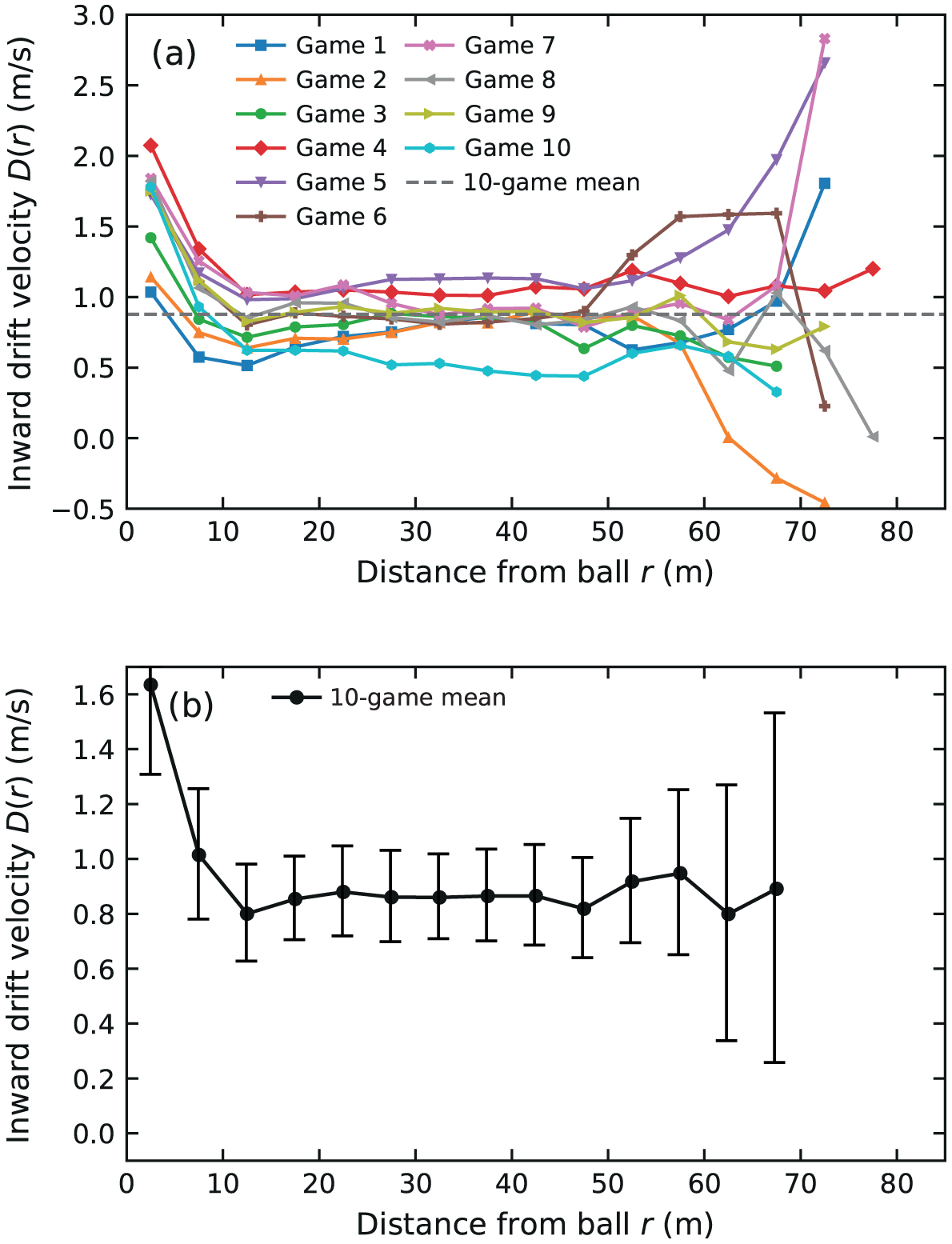}
\caption{(a)~Distance-dependent radial drift velocity $D(r)$ for the ten sample games. Positive values correspond to motion toward the ball. The horizontal dashed line indicates the equal-weight ten-game mean value, $0.89~\text{m}\,\text{s}^{-1}$. (b)~Average of the ten curves in (a). The error bars denote one standard deviation across the ten game-specific values of $D(r)$ in each distance interval. As a plotting convention, the points are placed at the centers of the 5-m distance intervals, with the first point at $r=2.5$~m.}
\label{fig:radial_drift}
\end{figure}

Figure~\ref{fig:radial_drift}(a) shows $D(r)$ for the ten games. In all cases the drift velocity is positive over almost all plotted distances, except for Game~2 at the largest distances, beyond about $65~\text{m}$. Thus, after removing the nearest player to the ball, the remaining
players, excluding the goalkeepers, still move on average toward the
instantaneous ball position. The equal-weight ten-game mean inward drift
velocity is
\begin{equation}
D_{\text{mean}} = 0.89 \pm 0.16~\text{m}\,\text{s}^{-1},
\label{eq:mean_drift}
\end{equation}
where the uncertainty is the standard deviation across the ten games. Comparing this velocity with the total player speeds plotted by
Novillo et al.~\cite{Novillo2024} as functions of player--ball distance,
their role-averaged outfield speeds are about
$2.4~\mathrm{m\,s^{-1}}$, whereas goalkeeper speeds are about
$1.1~\mathrm{m\,s^{-1}}$. The mean inward drift velocity in Eq.~\eqref{eq:mean_drift} is
about $40\%$ of a typical outfield player's total speed.
A value of $0.89~\mathrm{m\,s^{-1}}$ is comparable to a slow walking
speed.

Although the ten games differ in detail, the radial dependence has the
same overall form up to about $50~\mathrm{m}$. We therefore use the ten-game mean in
Fig.~\ref{fig:radial_drift}(b) to describe the common radial behavior.
As $r$ increases from the shortest retained separations, the mean drift
velocity decreases sharply to a shallow local minimum around
$r\simeq12~\mathrm{m}$ and then rises to a broad plateau of about
$0.9~\mathrm{m\,s^{-1}}$ extending to about $65~\mathrm{m}$. The error bars denote one standard deviation across
the ten game-specific values of $D(r)$ in each distance interval and
widen noticeably beyond about $50~\mathrm{m}$ as the data become
sparser. At the largest separations, the annular sampling region around
the ball is also increasingly truncated by the finite pitch. A
plausible explanation for the short-distance enhancement is that, even
after the nearest player has been excluded, other players only a few
meters from the ball are often directly involved in the local play,
through pressing, challenging, or support movements that have a large
component toward the ball. The broad plateau, in contrast, is a
role-averaged feature rather than evidence that all players have the
same distance dependence. The role dependence of $D(r)$ is examined in
Sec.~\ref{role}.

We compare the short-distance behavior with the speed--distance analysis
of Novillo et al.~\cite{Novillo2024}. They analyzed total player speed
and movement direction as functions of player--ball distance, player
role, and attacking or defensive phase. They found that total speed
decreases within approximately $2~\mathrm{m}$ of the ball, which they
attributed to close ball control. This is a different regime from the
short-distance enhancement observed here, because separations below
$2~\mathrm{m}$ are excluded from the present analysis. Novillo et al.
also averaged speed and movement direction separately, whereas $D(r)$
averages the signed inward component of each player's velocity. These
averages therefore cannot in general reconstruct
$D(r)$.

The game contexts and score histories are summarized in Table~\ref{tab:data_summary}. The ten games span a range of outcomes and scoring patterns, providing a useful test of whether the observed radial drift velocity persists under different match situations. Despite this variation, Fig.~\ref{fig:radial_drift}(a) shows the same broad behavior across all ten games. This supports the interpretation of $D(r)$ as a robust first
characterization of the distance dependence of the ball-directed drift
velocity, rather than as a feature of one particular game, outcome, or
score history.

\section{Ball position drift velocity field across the pitch}
\label{sec:pitch_field}

The radial average $D(r)$ is the distance law: the mean inward drift velocity at fixed player--ball distance. The field $K(x,y)$ is the complementary pitch-position average: the mean inward drift velocity when the ball is in a particular part of the pitch. In this second average, the player--ball distance $r$ is not fixed. The result is a position-dependent velocity: positive values mean average motion toward the ball, and negative values mean average motion away from it.

\begin{figure}[t]
\centering
\includegraphics[width=\linewidth]{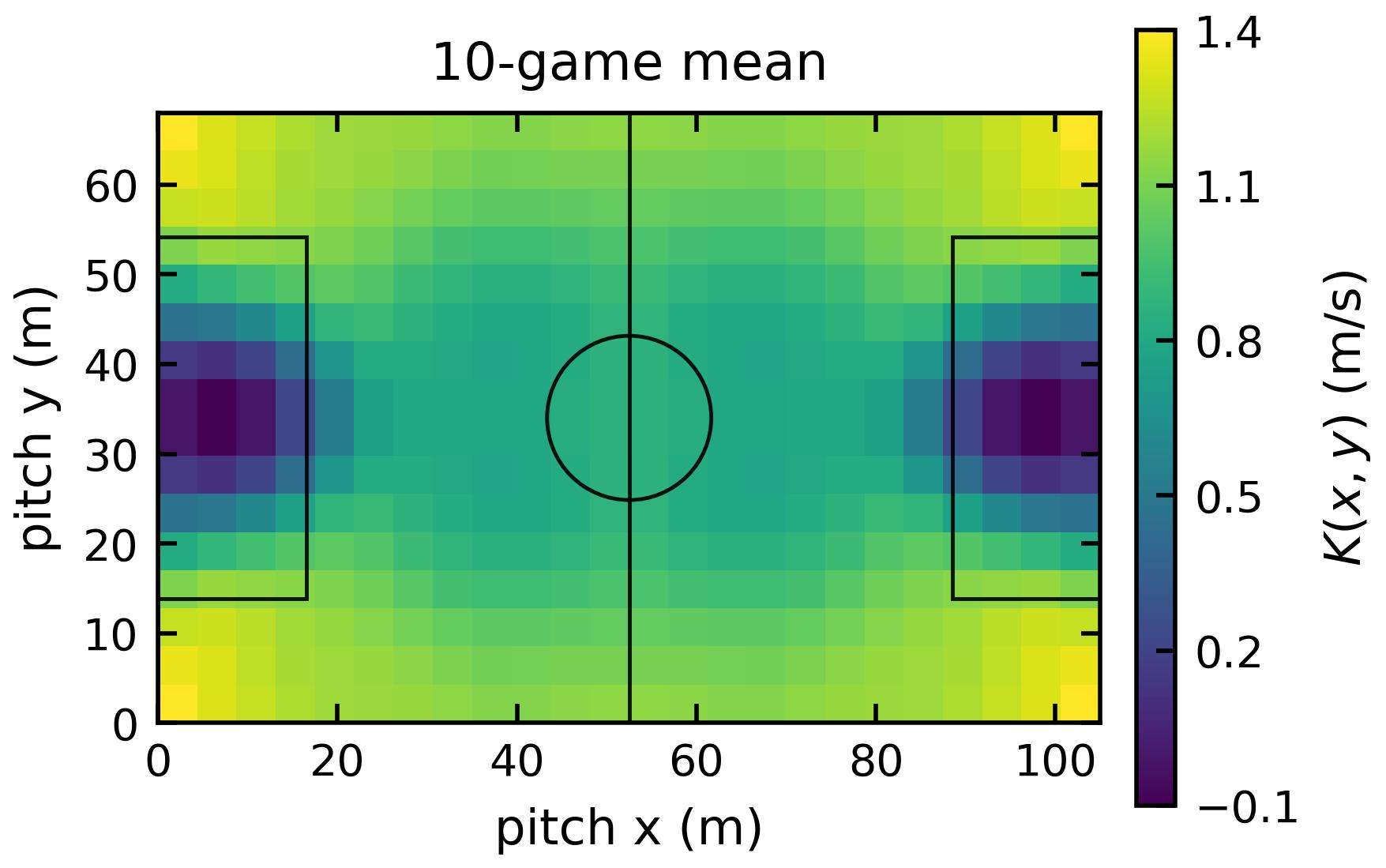}
\caption{
Symmetry-averaged ten-game mean map of $K(x,y)$ over the pitch. For each game, values at locations related by left--right and end-to-end pitch symmetry are combined, and the resulting field is subjected to count-weighted Gaussian smoothing with a standard deviation of one grid cell ($\sigma_x=4.375$~m and $\sigma_y=4.25$~m). The ten smoothed game maps are then averaged with equal weight. The color scale gives $K(x,y)$ in $\mathrm{m\,s^{-1}}$, i.e., the mean inward velocity component of the included players when the ball lies in the corresponding pitch cell. The pitch markings are included only to identify the spatial regions.
}
\label{fig:K_maps}
\end{figure}

Figure~\ref{fig:K_maps} shows the ten-game mean symmetry-averaged
$K(x,y)$ field. For each game, values at locations related by
left--right and end-to-end pitch symmetry are first averaged together.
The resulting symmetry-averaged field is then subjected to count-weighted
Gaussian smoothing with a standard deviation of one grid cell,
corresponding to $\sigma_x=4.375$~m and $\sigma_y=4.25$~m.
Specifically, the Gaussian filter is applied separately to the
accumulated inward-velocity sum and the number of observations in each
cell, and the smoothed field is obtained from their ratio. The ten
symmetry-averaged and smoothed fields are then averaged with equal
weight and displayed over the full pitch. This improves the statistics
and emphasizes the common pitch-scale structure. The corresponding
unsymmetrized $K(x,y)$ maps, obtained without the symmetry averaging but
with the same count-weighted smoothing, are shown in
Appendix~\ref{app:unfolded_team}.

The values of $K(x,y)$ are directly comparable with the values of $D(r)$,
because both are averages of the same inward velocity $D_i(t)$. In the
absence of smoothing, averaging the cell values of $K(x,y)$ over the pitch
with weights $N_{xy}$, the number of retained $D_i(t)$ values entering
Eq.~\eqref{eq:K_field} in each cell, gives for each game the same global
mean inward drift velocity as averaging $D_i(t)$ directly. Taking the equal-weight
mean of these game-wise values then gives
$D_{\rm mean}=0.89~\mathrm{m\,s^{-1}}$, up to rounding.
Figure~\ref{fig:K_maps} should therefore be read as a smoothed spatial
decomposition of this mean: most regions are close to the global mean,
whereas the goalmouth regions lie well below it.

The figure shows several recognizable characteristics. Away from the goalmouths, $K(x,y)$ varies only weakly across the main part of the pitch. The attraction toward the ball is not produced by one special part of the pitch, but is present over a wide range of ball positions.

Near the goalmouths, $K(x,y)$ is strongly reduced and becomes slightly
negative in well-defined regions. In the effective-field analogy, these
regions correspond to weak local repulsion. When the ball is
close to goal, some of the included players hold position, move away from
the ball to recover shape, or prepare for the next phase of play. Because
the player nearest to the ball has been omitted, $K(x,y)$ in these regions
mainly reflects the average motion of the remaining players.

There is no contradiction between the local negative values of $K(x,y)$ near the goalmouths and the positive radial curve $D(r)$. The two quantities average the same set of individual drift velocities in different ways: in the calculation of $D(r)$, the average is taken over all ball positions at fixed player--ball distance, whereas in the calculation of $K(x,y)$, the average is taken over all retained times for which the ball lies in a given pitch cell. Local goalmouth behavior can appear clearly in $K(x,y)$ while being averaged together with the rest of the pitch in $D(r)$.

\begin{figure}[t]
\centering
\includegraphics[width=\linewidth]{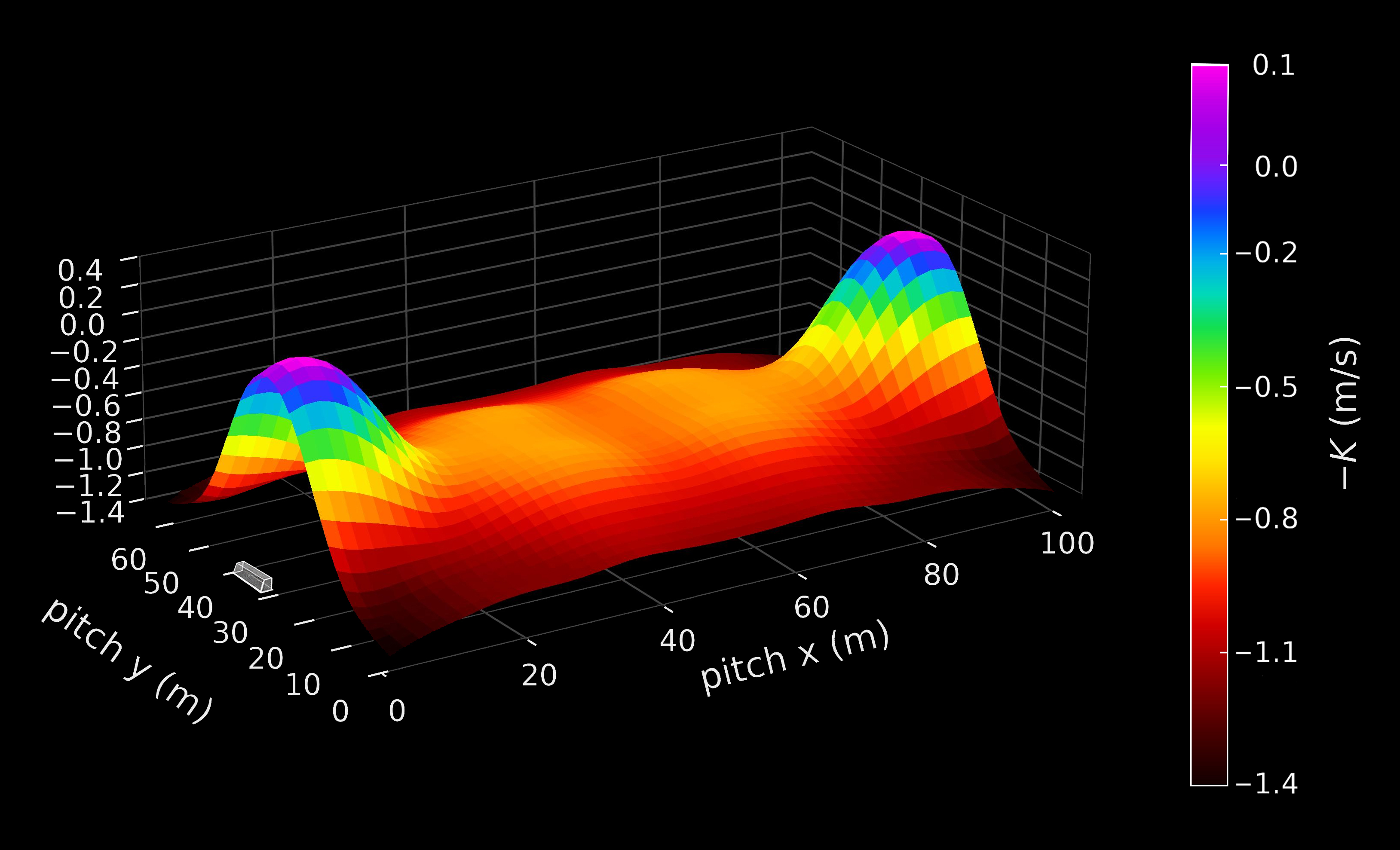}
\caption{Three-dimensional visualization of the ten-game mean
$K(x,y)$ map in Fig.~\ref{fig:K_maps}. The plotted surface is
$-K(x,y)$, so reduced values of $K(x,y)$ appear as raised features.}
\label{fig:K_surface}
\end{figure}

Figure~\ref{fig:K_surface} shows the same ten-game mean field as Fig.~\ref{fig:K_maps}, but plotted as $-K(x,y)$. With this convention, regions of reduced inward drift velocity appear as raised features, making the goalmouth suppression especially clear.

A second, weaker feature is visible near the center of the pitch as a broad region of reduced $K(x,y)$, which appears as a shallow elevation in the $-K(x,y)$ representation of Fig.~\ref{fig:K_surface}. For quantitative analysis, $K(x,y)$ was averaged over an approximately $60\times30$~m region centered on the pitch midpoint.
 The value used for comparison was the mean of $K(x,y)$ over the much larger part of the map excluding the goalmouths, rather than over only the cells immediately surrounding the central region. On the $105 \times 68~\mathrm{m}$ pitch, let $x_g=\min(x,105-x)$ be the distance from the nearest goal line, and let $y_c=|y-34|$ be the distance from the pitch centerline. Cells with $x_g<13~\mathrm{m}$ and $y_c<14~\mathrm{m}$ were excluded from the non-goalmouth average. This removes the central region in front of either goal, extending to about $\sqrt{13^2+14^2}\simeq19~\mathrm{m}$ from the center of each goal. With this definition, the 10-game mean gives $K_{\mathrm{non-goal}}\simeq0.99~\mathrm{m\,s^{-1}}$, whereas the central region has $K_{\mathrm{center}}\simeq0.84~\mathrm{m\,s^{-1}}$. The mean inward drift velocity in the central region is therefore lower
than the non-goalmouth average by
about $15\%$.

The central depression should not be interpreted simply as evidence that teams prefer attacking down the wings. Because the nearest player to the ball is excluded, the average represented by $K(x,y)$ is mainly the motion of the remaining players. When the ball is central, one player is often already engaging the ball carrier, whereas the other players have different tasks: defenders mark receivers, screen passing lanes, protect both sides of the pitch, or hold the defensive line, and attackers move away from the ball to create passing options across the pitch or closer to the opponent's goal. 
These useful movements need not have a large component toward the ball. If too many players moved rapidly toward a central ball, they would open side lanes and space behind them. Wide ball positions are different: play is compressed against a boundary, and many recovery, pressing, and support runs have a larger component toward the ball. The appearance of a central depression in $K(x,y)$ therefore implies that, for central ball positions, the remaining players have a weaker average inward drift velocity, not that the central region is unimportant. The role dependence of $K(x,y)$ is examined in
Sec.~\ref{role}.

A natural question is whether the player density could be responsible for the central depression. To test this, a player-density map was constructed from the same retained periods of play used for the drift velocity analysis, as shown in Appendix~\ref{app:player_density}. The nearest player to the ball was included because the purpose here is only to count player positions. At the resolution of the grid, the largest player density occurs at the center of the pitch. The central depression in $K(x,y)$ is not caused by a lack of players in the central region. Players are often concentrated there, but their average motion has a smaller component toward the ball.

To sum up, $D(r)$ and $K(x,y)$ provide complementary averages of the same radial drift velocity: $D(r)$ describes its dependence on player--ball distance, whereas $K(x,y)$ shows how the inward drift velocity varies with the ball's position
on the pitch.

\section{Dependence on player role}
\label{role}

The averages considered so far combine outfield players with different
playing roles. There is no reason to expect defenders, midfielders, and
attackers to contribute in the same way to the inward drift velocity. Indeed, Novillo et al.~\cite{Novillo2024} found pronounced
role-dependent differences in both total player speed and the direction of
motion relative to the ball. We therefore repeat the present analysis
separately for the registered attacker, midfielder, and defender roles.
Reliable role labels are available for Games~3--10, so this part of the
analysis uses those eight games. Their broad positional composition is
summarized in Table~\ref{tab:role_composition}. For each role, $D(r)$ and
$K(x,y)$ are first calculated separately for each game and are then averaged
with equal weight over the eight games. All other analysis choices remain
unchanged.

\begin{table}[t]
\caption{Broad positional composition for Games~3--10. The entries give the
numbers of attackers (ATT), midfielders (MID), and defenders (DEF), summed
over the two teams and excluding the two goalkeepers. At this three-role
level the numbers are constant within each match, including across
substitutions.}
\label{tab:role_composition}
\begin{ruledtabular}
\begin{tabular}{lc}
Game & ATT / MID / DEF \\
\hline
Game 3  & $3/9/8$  \\
Game 4  & $2/10/8$ \\
Game 5  & $2/10/8$ \\
Game 6  & $4/8/8$  \\
Game 7  & $3/10/7$ \\
Game 8  & $3/9/8$  \\
Game 9  & $4/10/6$ \\
Game 10 & $3/9/8$  \\
\end{tabular}
\end{ruledtabular}
\end{table}

\begin{figure}[t]
\centering
\includegraphics[width=\linewidth]{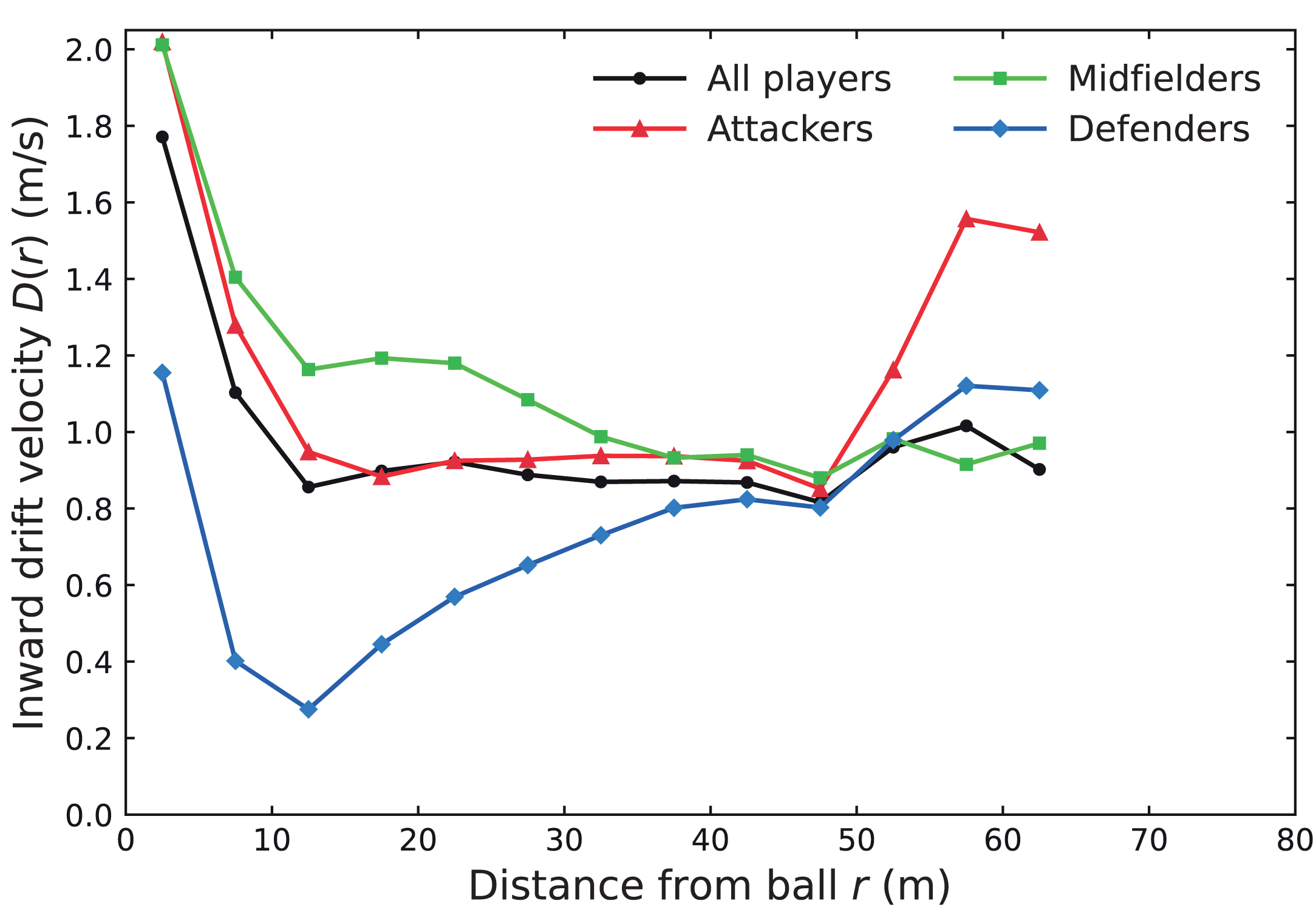}
\caption{Mean radial drift velocity separated by player role. The attacker,
midfielder, and defender curves are equal-weight averages over
Games~3--10, for which registered outfield-player roles are available.
For comparison, the solid black curve is the corresponding eight-game mean
without role separation. Only $5$-m player--ball distance intervals to
which all eight games contribute are shown. Goalkeepers and the nearest
player are excluded. Positive values denote mean
motion toward the ball.}
\label{fig:role_Dr}
\end{figure}

The radial curves show a clear dependence on player role. As shown in
Fig.~\ref{fig:role_Dr}, attackers and midfielders have the strongest
inward drift velocity at low $r$, both reaching about
$2.0~\mathrm{m\,s^{-1}}$. The defender curve is lower than either, and it
decreases to a pronounced minimum near $r\simeq12~\mathrm{m}$ before
increasing with distance. The midfielder curve decreases gradually with
increasing $r$, whereas the attacker curve remains close to
$0.9~\mathrm{m\,s^{-1}}$ over roughly $15$--$50~\mathrm{m}$ and then
increases at larger $r$. The three role curves become most similar over
about $35$--$50~\mathrm{m}$. We discuss these curves in more detail below.

The shallow local minimum near $r\simeq12~\mathrm{m}$ noted earlier in
the role-averaged $D(r)$ is not accidental: it arises from the
pronounced minimum in the defender contribution. A plausible explanation
is tactical. At player--ball separations of order $10$--$15~\mathrm{m}$,
defenders may hold the defensive line, mark opponents, or move laterally
to preserve shape rather than close directly on the ball. Such motion
reduces the inward component of their velocity without necessarily
reducing their total speed.

At large $r$, the attacker curve develops a pronounced maximum near
$r\simeq60~\mathrm{m}$. The mean attacker inward drift velocity is about
$0.9~\mathrm{m\,s^{-1}}$ over $10$--$50~\mathrm{m}$ and about
$1.4~\mathrm{m\,s^{-1}}$ over $50$--$65~\mathrm{m}$, with game-to-game
standard deviations of about $0.2~\mathrm{m\,s^{-1}}$. In all eight games,
the latter mean is higher than the former. (The estimated probability of
obtaining so consistent a difference by chance is $0.8\%$.) A weaker
local maximum visible in the overall $D(r)$ around $50$--$60~\mathrm{m}$
probably reflects the same effect, diluted because attackers account for
only $2$--$4$ of the $20$ outfield positions in these games. One possible
explanation for the attacker maximum is that such large separations occur
when play has moved far from the advanced attackers, who then make long
runs toward the ball to rejoin the play.

The smoother midfielder curve may reflect their role in connecting defense and attack. Midfielders often move with the play and provide support around the ball, so their motion can retain a substantial inward component over a broad range of $r$. This role may also explain why their curve lacks both the small-$r$ defender minimum and the large-$r$ attacker maximum. As $r$ increases, their motion may more often involve maintaining position and team shape rather than moving directly toward the ball, leading to a gradual reduction in the inward component.

The role-resolved results of Novillo et al.~\cite{Novillo2024} provide
a useful comparison. Their total-speed curves vary comparatively
smoothly with player--ball distance, with no counterpart of the
pronounced defender minimum near $r\simeq12~\mathrm{m}$, and their
continuous analysis ends at $50~\mathrm{m}$, before the attacker maximum
found here. The role-dependent structure in $D(r)$ therefore provides
information complementary to their role-resolved speed analysis.

\begin{figure}[t]
\centering
\includegraphics[width=\linewidth]{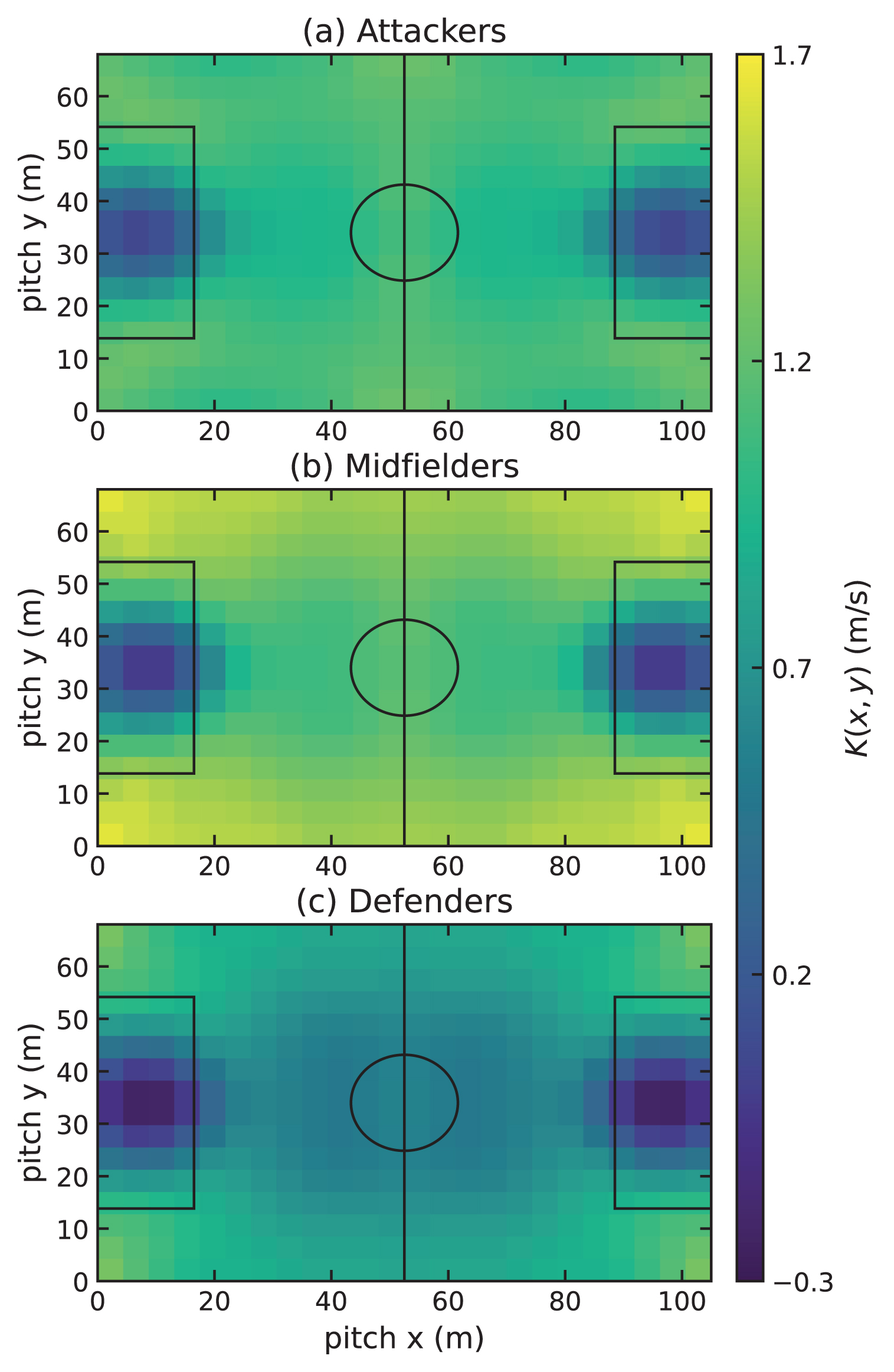}
\caption{Mean spatial drift-velocity field separated by player role:
(a) attackers, (b) midfielders, and (c) defenders. Each map is the
equal-weight average over Games~3--10 and uses the same $24\times16$
pitch grid, symmetry averaging, and count-weighted Gaussian smoothing
as the main $K(x,y)$ map. The common color scale runs from
$-0.3$ to $1.7~\mathrm{m\,s^{-1}}$. Positive values denote mean motion
toward the ball and negative values mean motion away from it.}
\label{fig:role_K}
\end{figure}

The spatial fields can be used to show how this role dependence varies with ball
position. Figure~\ref{fig:role_K} shows the mean $K(x,y)$ fields for
the same three roles. All three have reduced inward drift velocity near the
goalmouths, but the magnitude and sign depend on role.

In the smoothed eight-game mean fields, $K(x,y)$ remains non-negative
throughout for attackers and midfielders, with minimum values of about
$0.07~\mathrm{m\,s^{-1}}$ and effectively zero, respectively. Only
the defenders show negative values near the goals, reaching about
$-0.2~\mathrm{m\,s^{-1}}$. Thus the effective repulsion near the
goalmouths in the combined outfield-player $K(x,y)$ field arises
entirely from defenders: attackers and midfielders continue to move,
on average, toward the ball there.

Analysis of the fields in Fig.~\ref{fig:role_K} shows that the central
depression discussed in Sec.~\ref{sec:pitch_field} is present for all three
roles and is strongly role dependent. Using the same central-region and
non-goalmouth definitions, the reduction is about $5\%$ for attackers,
$11\%$ for midfielders, and $22\%$ for defenders. This is consistent
with the earlier interpretation that, for central ball positions,
tactically useful movements by the surrounding players need not have a
large component toward the ball.

The appendices give the numerical details and robustness tests behind the results presented above. Appendix~\ref{app:filtering} contains the filtering, binning, and data volumes used in the calculations. Appendix~\ref{app:nearest_player_check} gives the nearest-player inclusion test. Appendix~\ref{app:unfolded_team} presents the $K(x,y)$ maps before symmetry averaging and reports the team-by-team comparison. Appendix~\ref{app:player_density} gives the player-density test discussed above in connection with the central depression. Appendix~\ref{app:goalkeeper_half} contains goalkeeper-inclusion and half-match tests, together with goalkeeper-resolved results for $D(r)$ and $K(x,y)$.

\section{Discussion and conclusions}
\label{sec:discussion}

Using tracking data from ten games, we provide a quantitative
description of a familiar feature of football: players reorganize their
motion around the ball. The extracted drift velocity is not a force or an
acceleration, but the signed component of each player's velocity along the
direction toward the instantaneous ball position. By excluding restarts
and constrained phases of play, the goalkeepers, the player nearest to the
ball, and very small player--ball separations, we focus on the
surrounding outfield players during open play.

The equal-weight mean inward drift velocity over the ten games is about
$0.9~\mathrm{m\,s^{-1}}$. This is not the players' total running speed, but
the component directed toward the ball after averaging over many
player--time observations. The mean drift velocity is positive in all ten
games, despite differences in match outcome, score history, and
drift-velocity magnitude.

The distance-dependent radial drift velocity $D(r)$ shows a pronounced
short-range enhancement followed by a broad plateau. At the shortest
analyzed separations, $D(r)$ rises to about $1.6~\mathrm{m\,s^{-1}}$ even
after the nearest player has been excluded, before decreasing rapidly and
passing through a shallow local minimum near $r\simeq12~\mathrm{m}$.
At larger distances, the ten-game mean remains close to its overall value
over a broad range. Thus the radial dependence consists of a short-range
enhancement, a shallow minimum, and an extended region of approximately
constant inward drift velocity rather than an inverse-power law.

The ball-position field $K(x,y)$ provides the complementary spatial
description. Away from the goalmouths, $K(x,y)$ is broadly positive
across the pitch. A broad central depression lies about $15\%$ below the
non-goalmouth average, whereas the reduction is stronger near the
goalmouths, where $K(x,y)$ can become slightly negative. In the
effective-field analogy, the predominantly attractive field 
develops weak local repulsion near the goals.

For the eight games with registered player roles, separating the players
into attackers, midfielders, and defenders reveals the origin of several
of these features. At short range, attackers and midfielders have the
strongest inward drift velocity, both reaching about $2~\mathrm{m\,s^{-1}}$,
whereas the defender curve falls to a much deeper minimum of about
$0.3~\mathrm{m\,s^{-1}}$ near $r\simeq12~\mathrm{m}$. This defender
minimum accounts for the much shallower minimum in the role-averaged
$D(r)$. The three curves become most similar at intermediate distances.
At larger distances, around 60~$\mathrm{m}$, the attacker inward drift velocity develops a
pronounced maximum, whereas the defender inward drift velocity continues
to increase and the midfielder curve remains relatively flat. The spatial
fields give an even clearer separation: $K(x,y)$ remains non-negative
for attackers and midfielders, whereas only defenders produce negative
values near the goals. Thus the weak effective repulsion near the
goalmouths arises entirely from defenders.

The method used to construct $D(r)$ and $K(x,y)$ is not specific to
association football. In any team sport in which players move in relation
to a ball, puck, or ball carrier, their velocities can be projected toward
that moving reference and averaged in the same way. In basketball,
handball, ice hockey, lacrosse, rugby, or American football, $D(r)$ could
quantify how strongly players move toward it as a function of distance,
whereas $K(x,y)$ could characterize how this tendency varies across the
court, rink, or field. 
For football itself, these drift-velocity measures could provide compact
descriptors of playing style. A larger $D(r)$ at a given distance would
indicate a stronger tendency to close on the ball, whereas a smaller
value would indicate that players more often maintain spacing, delay, or
move less directly toward it. From the complementary field $K(x,y)$, one can identify where on the
pitch these tendencies are strongest or weakest. Such
quantities should not by themselves be interpreted as measures of
performance, but they provide objective observables for comparing teams,
playing roles, match states, halves, score states, levels of play, and
changes over a season.

We deliberately limit the present study to ten public matches and to the
radial component of player motion relative to the instantaneous ball
position. We do not explicitly model possession, passing choices,
pressing triggers, defensive lines, tactical formations, or individual
decision making.
Nevertheless, the principal result is robust: over much of open play,
the collective motion of the surrounding players contains a persistent
inward component toward the ball, with a characteristic dependence on
player--ball distance, systematic spatial variation across the pitch,
and clear differences between playing roles. The ball is not a planet,
and football players are not satellites orbiting it, but the
tracking data nevertheless reveal a measurable effective field that
provides a compact statistical description of how player motion is
organized around the ball.

\begin{acknowledgments}
The author thanks Metrica Sports and Bassek \textit{et al.} for making the Metrica Sports and Bundesliga tracking data, respectively, publicly available.
\end{acknowledgments}

\appendix

\section{Filtering and numerical details}
\label{app:filtering}

This appendix gives the numerical filtering rules used to produce the retained data in Table~\ref{tab:appendix_data_volume}. Games~1 and 2 are supplied as standard Metrica CSV tracking and event files, whereas Game~3 is supplied in a later Metrica format with separate tracking, metadata, and event files. Games~4--10 use the Bundesliga data, with separate tracking, event, and match-information files. The same exclusion window was used for all ten games, although the markers used to define it differed between the data formats.

Let $t_0$ and $t_1$ denote the start and end times associated with one
such stoppage or constrained-play marker. For Games~1 and 2, these were
annotated restart or interruption events, including corners, throw-ins,
goal kicks, kickoffs, free kicks, fouls, cards, ball-out events, and
similar interruptions. For Game~3, the event file was not used for this
filtering step; instead, the markers were intervals in which the public
ball position was missing or outside the pitch. For Games~4--10, the
markers were derived from the Bundesliga event data. Each event provides
an event-time timestamp, and some events additionally provide a
decision-time timestamp. When both timestamps were present, $t_0$ was
taken as the earlier of the event time and decision time and $t_1$ as
the later. When no decision time was provided, $t_0=t_1$ was set equal
to the event time. In all cases, tracking frames satisfying
\begin{equation}
t_0 - 3~\mathrm{s} \leq t \leq t_1 + 10~\mathrm{s}
\label{eq:exclusion_window}
\end{equation}
were excluded.

After this time-window filtering, a valid ball coordinate was required, and the neighboring frames needed for the central-difference velocity in Eq.~\eqref{eq:velocity} had to be available within the same period. For the Bundesliga data, the supplied ball-status flag was also required to indicate that the ball was in play. At the player level, both goalkeepers were excluded, followed by the remaining player nearest to the ball at each frame, and observations with $r<2~\mathrm{m}$ were removed before forming the averages. Referee tracks, where present, were not included in the player set. The resulting data volumes are shown in Table~\ref{tab:appendix_data_volume}. The derived stoppage-window files used to define the retained data are provided in the Supplemental Material~\cite{supplemental}. This contains the ten game-specific exclusion-window files, a game-to-file mapping, and a description of the marker definitions and common exclusion procedure.

\begin{table*}[t]
\caption{Data volumes after filtering for the 10 games. The penultimate
column gives the number of individual values of $D_i(t)$ entering the
averages reported in the main text after all exclusions described. Each value
corresponds to one retained player at one retained time; the final
column gives the corresponding per-game mean $D_i(t)$.}
\label{tab:appendix_data_volume}
\begin{ruledtabular}
\begin{tabular}{lccccc}
Game & Raw time ($\mathrm{min}$) & Included time ($\mathrm{min}$) & Included frames & Values of $D_i(t)$ & Mean $D_i(t)$ ($\mathrm{m\,s^{-1}}$) \\
\hline
Game 1  & $96.67$ & $39.95$ & $59{,}924$ & $1.133\times10^6$ & $0.707$ \\
Game 2  & $94.10$ & $37.75$ & $56{,}620$ & $1.069\times10^6$ & $0.756$ \\
Game 3  & $95.84$ & $35.55$ & $53{,}324$ & $1.009\times10^6$ & $0.823$ \\
Game 4  & $97.31$ & $38.31$ & $57{,}472$ & $1.086\times10^6$ & $1.095$ \\
Game 5  & $91.48$ & $30.31$ & $45{,}472$ & $0.859\times10^6$ & $1.104$ \\
Game 6  & $94.37$ & $26.72$ & $40{,}075$ & $0.719\times10^6$ & $0.917$ \\
Game 7  & $95.02$ & $32.02$ & $48{,}026$ & $0.908\times10^6$ & $1.020$ \\
Game 8  & $97.47$ & $36.84$ & $55{,}266$ & $1.045\times10^6$ & $0.921$ \\
Game 9  & $94.90$ & $40.61$ & $60{,}911$ & $1.138\times10^6$ & $0.932$ \\
Game 10 & $97.87$ & $41.56$ & $62{,}344$ & $1.180\times10^6$ & $0.609$ \\
\hline
Total & $955.04$ & $359.62$ & $539{,}434$ & $10.146\times10^6$ & $-$ \\
\end{tabular}
\end{ruledtabular}
\end{table*}

\begin{figure*}[t]
\centering
\includegraphics[width=0.8\linewidth]{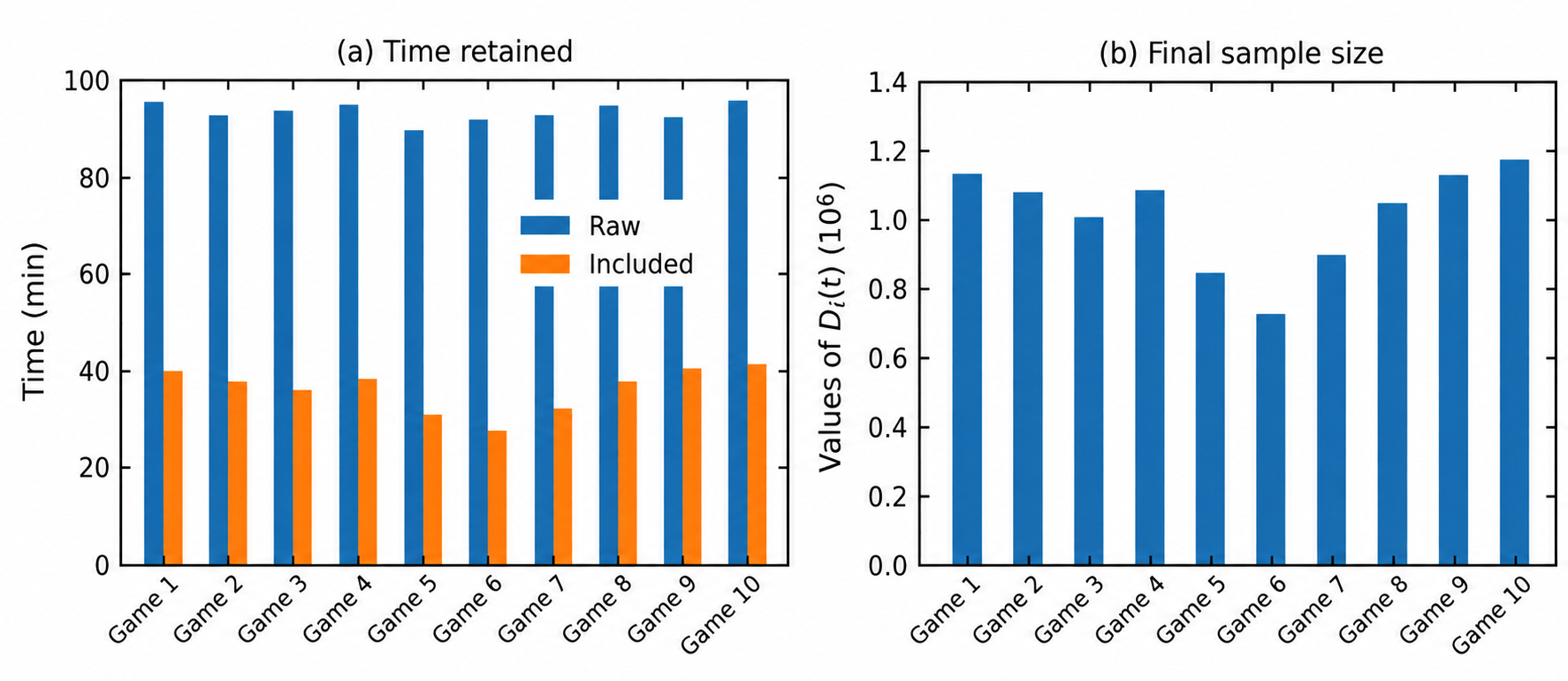}
\caption{Graphical summary of the filtering in Table~\ref{tab:appendix_data_volume}. (a) Raw tracked time and included open-play time for the ten games. (b) Final number of retained values of $D_i(t)$ after all exclusions.}
\label{fig:appendix_filtering_summary}
\end{figure*}

The retained times and the final number of $D_i(t)$ values for each game are shown graphically in Fig.~\ref{fig:appendix_filtering_summary}. Although the filtering removes a substantial part of each match, the retained data are comparable in size across the ten games.

\section{Nearest-player inclusion test}
\label{app:nearest_player_check}

In the analysis presented in the main text, the player nearest to the
ball is omitted at each included frame. This should not be interpreted as an attempt to identify possession. The rule is purely kinematic: the omitted player is simply the player with the smallest distance to the ball in that frame. In many cases this player will be carrying, receiving, shielding, or challenging for the ball, but the rule does not require a possession label. Its purpose is to remove the player most directly constrained by the ball and to analyze the motion of the surrounding players.

It is useful to test explicitly whether this choice changes the calculated averages. The expectation is different for $D(r)$ and for $K(x,y)$. For the radial curve, including the nearest player can add one extra observation per retained frame, provided that the observation satisfies the same player-level validity conditions and $r\ge2$~m cutoff. When added, that observation belongs to the distance interval containing the minimum player--ball distance in that frame. Therefore it should mainly affect the shortest-distance intervals and should not materially change the intermediate- or large-distance part of $D(r)$, except in rare frames where even the nearest player is far from the ball.

For the field $K(x,y)$, the values entering the average are determined by the pitch cell in which the ball lies, rather than by the player--ball distance. Let $K_{\mathrm{excl}}(x,y)$ denote the cell average used in the analysis presented in the main text, with the nearest player excluded, and let $K_{\mathrm{incl}}(x,y)$ denote the corresponding cell average obtained when that player is included. Let $M_{\mathrm{add}}(x,y)$ denote the number of retained frames in which the ball lies in the cell and the nearest-player observation also satisfies the $r\geq2$ m cutoff, so that it is actually added to the average. Let $k_{\mathrm{near}}(x,y)$ denote the mean radial drift velocity of these added nearest-player observations. If $N_{\mathrm{excl}}(x,y)$ is the number of retained non-nearest-player observations in the cell, then $\bar{N}(x,y)\equiv N_{\mathrm{excl}}(x,y)/M_{\mathrm{add}}(x,y)$ is the ratio of the number of non-nearest-player observations already contributing to the cell average to the number of nearest-player observations that are added. For cells with $M_{\mathrm{add}}(x,y)>0$, including the nearest player changes the cell average according to
\begin{equation}
\begin{aligned}
K_{\mathrm{incl}}(x,y)-K_{\mathrm{excl}}(x,y)
&=
\frac{k_{\mathrm{near}}(x,y)-K_{\mathrm{excl}}(x,y)}
{\bar{N}(x,y)+1}.
\end{aligned}
\label{eq:nearest_player_shift}
\end{equation}
For any cell with $M_{\mathrm{add}}(x,y)=0$, no nearest-player observation is added and $K_{\mathrm{incl}}(x,y)=K_{\mathrm{excl}}(x,y)$. Thus $K(x,y)$ can shift when the nearest player is included, but the effect is diluted by the much larger number of retained non-nearest-player observations already contributing to each cell average. For the 10-game average, including the nearest player changes the arithmetic mean of $K(x,y)$ over the 384 pitch cells from $0.918$ to $0.927~\mathrm{m\,s^{-1}}$. The maximum absolute difference between $K_{\rm incl}(x,y)$ and $K_{\rm excl}(x,y)$ in any cell is $0.041~\mathrm{m\,s^{-1}}$, although the corresponding fractional difference can be appreciable in cells where $K(x,y)$ is close to zero. These results are consistent with the cell-by-cell dilution expressed by
Eq.~\eqref{eq:nearest_player_shift}.

\begin{figure*}[t]
\centering
\includegraphics[width=0.75\linewidth]{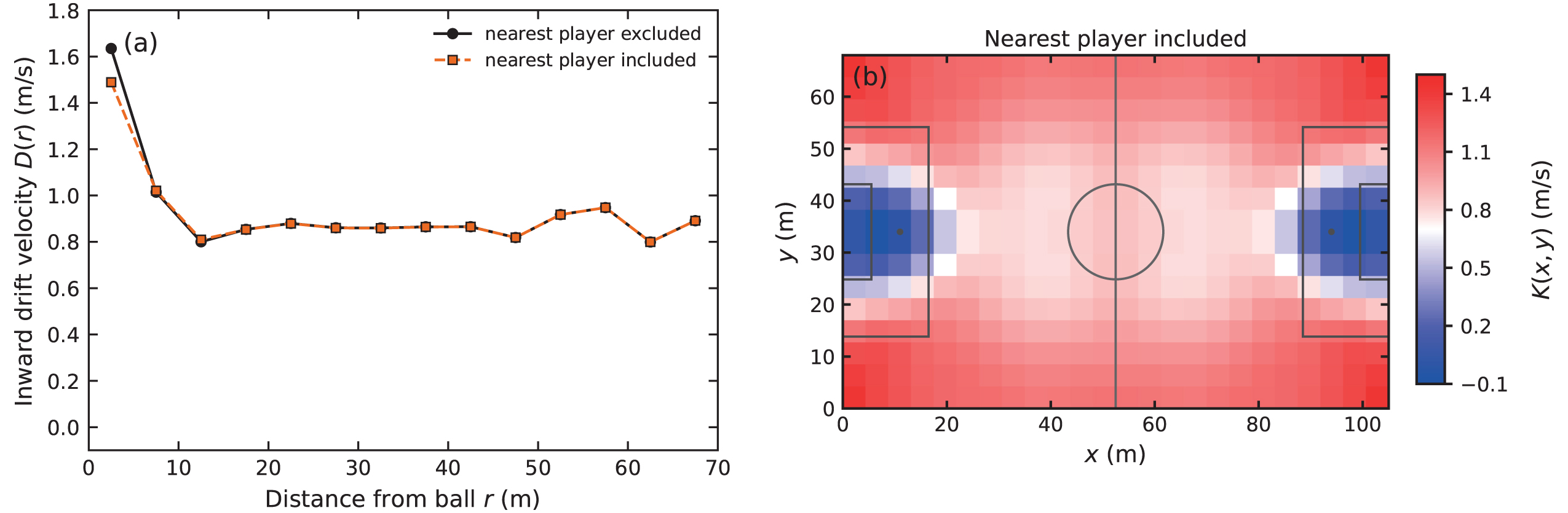}
\caption{Nearest-player inclusion test.
(a) Ten-game mean radial drift velocity curve $D(r)$ with the nearest player
excluded (solid black) and included (dashed orange), with all other
analysis choices unchanged.
(b) Ten-game symmetry-averaged $K(x,y)$ map with the nearest player
included.}
\label{fig:appendix_nearest_player_check}
\end{figure*}

Figure~\ref{fig:appendix_nearest_player_check} shows the numerical
effect of restoring the nearest player. In panel~(a), the solid black curve, labeled ``nearest player excluded,'' is the
ten-game mean $D(r)$ shown in Fig.~\ref{fig:radial_drift}(b). The orange dashed curve is a separate robustness calculation in which
the nearest player is restored while all other analysis choices are
left unchanged. Restoring this player lowers the first-bin mean, but has almost no effect at larger
player--ball distances. There is no contradiction between the two
short-distance values: they correspond to two different player
populations. Adding the nearest player lowers the first-bin average
because the mean $D_i(t)$ of the added observations is below the mean
of the observations already present in that interval.

Panel~(b) shows the corresponding ten-game symmetry-averaged $K(x,y)$
field when the nearest player is restored. The spatial pattern changes
very little. The Pearson correlation with the corresponding map in Fig.~\ref{fig:K_maps},
obtained with the nearest player excluded, is $0.9996$, and the mean absolute
cell-by-cell difference is only about $0.009~\mathrm{m\,s^{-1}}$. This
small change is consistent with Eq.~\eqref{eq:nearest_player_shift},
because at most one additional player observation is added per retained
frame to the many player observations contributing to each pitch cell.

The positive inward drift velocity is therefore not an artifact of excluding the
nearest player. The exclusion defines the population considered in the main text: the
motion of the surrounding outfield players rather than that of the player
most directly constrained by the ball. Restoring
that player produces a visible reduction only in the shortest-distance
part of $D(r)$ and leaves both the plateau at larger $r$ and the spatial
structure of $K(x,y)$ essentially unchanged.

\section{Full-pitch $K(x,y)$ maps and team-by-team comparison}
\label{app:unfolded_team}

\begin{figure*}[t]
\centering
\includegraphics[width=1.0\linewidth]{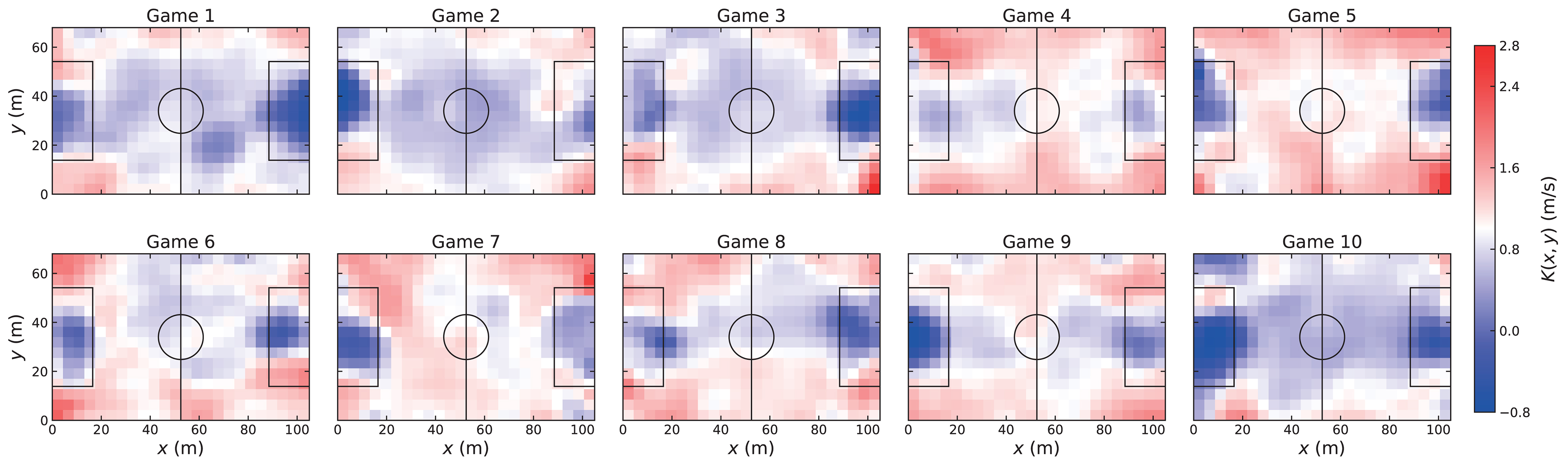}
\caption{Full-pitch maps of $K(x,y)$ before symmetry averaging.
Each panel shows one game on the original pitch coordinates after the
same filtering and count-weighted Gaussian smoothing used for the
symmetry-averaged $K(x,y)$ maps, with a Gaussian standard deviation
of one grid cell ($\sigma_x=4.375$~m and $\sigma_y=4.25$~m).}
\label{fig:appendix_unfolded_maps}
\end{figure*}

The $K(x,y)$ maps in the main text are symmetry-averaged to emphasize
the common structure of the measured fields. Figure~\ref{fig:appendix_unfolded_maps} shows the corresponding full-pitch, game-resolved $K(x,y)$ fields before symmetry averaging. The purpose is to make explicit what is simplified by the symmetry operation. The individual games contain end-to-end and left--right asymmetries, as expected for real matches, but the broad features of the field remain visible before folding.

As a second comparison, the same calculation was repeated separately for
the two teams in each game. The values of $D_i(t)$ were split by team,
and both $D(r)$ and $K(x,y)$ were recomputed from the two subsets.
Table~\ref{tab:team_appendix} summarizes the results. The team mean drift velocities differ by
no more than $26\%$ in the present sample. The team-resolved $D(r)$
curves have Pearson correlations between $0.29$ and $0.97$.

\begin{figure*}[t]
\centering
\includegraphics[width=1.0\linewidth]{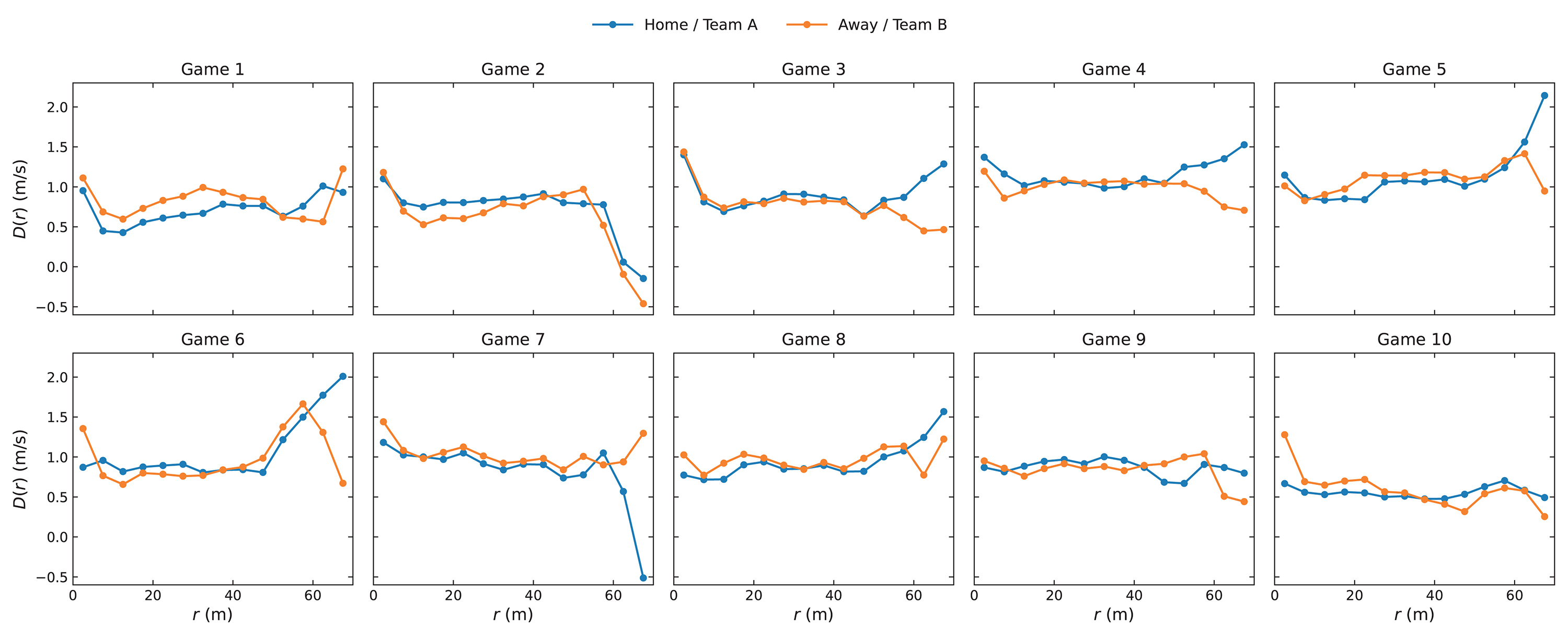}
\caption{Team-separated radial drift velocity curves. The values of $D_i(t)$
were split by team before constructing $D(r)$. For Game~3, the two
sides are labeled Team~A and Team~B.}
\label{fig:appendix_team_radial}
\end{figure*}

\begin{table*}[t]
\caption{Team-by-team comparison for the ten games. Team~1 and Team~2 follow the order of the two sides in Table~\ref{tab:data_summary}. The percentage difference is $100\times|\bar D_1-\bar D_2|/\bar D$, where $\bar D$ is the mean of the two team mean drift velocities. Correlations (labeled ``corr.'') are Pearson correlations between corresponding sampled distance intervals or corresponding symmetry-averaged $K(x,y)$ grid cells. The $D(r)$ correlations use distance intervals for which both team curves contain at least $10^3$ values. All calculations use the same analysis choices as in the main text.}
\label{tab:team_appendix}
\begin{ruledtabular}
\begin{tabular}{lccccc}
Game & \begin{tabular}[c]{@{}c@{}}Team 1\\ mean $D_i(t)$ ($\mathrm{m\,s^{-1}}$)\end{tabular} & \begin{tabular}[c]{@{}c@{}}Team 2\\ mean $D_i(t)$ ($\mathrm{m\,s^{-1}}$)\end{tabular} & \% Difference & $D(r)$ corr. & $K(x,y)$ corr. \\
\hline
Game 1 & $0.615$ & $0.799$ & $26.0\%$ & $0.29$ & $0.981$ \\
Game 2 & $0.820$ & $0.693$ & $16.7\%$ & $0.90$ & $0.953$ \\
Game 3 & $0.830$ & $0.816$ & $1.6\%$ & $0.91$ & $0.975$ \\
Game 4 & $1.129$ & $1.060$ & $6.3\%$ & $0.83$ & $0.959$ \\
Game 5 & $1.056$ & $1.151$ & $8.7\%$ & $0.92$ & $0.988$ \\
Game 6 & $0.902$ & $0.934$ & $3.5\%$ & $0.96$ & $0.985$ \\
Game 7 & $1.009$ & $1.030$ & $2.0\%$ & $0.90$ & $0.960$ \\
Game 8 & $0.932$ & $0.911$ & $2.3\%$ & $0.91$ & $0.904$ \\
Game 9 & $0.923$ & $0.942$ & $2.1\%$ & $0.84$ & $0.988$ \\
Game 10 & $0.626$ & $0.593$ & $5.4\%$ & $0.97$ & $0.972$ \\
\end{tabular}
\end{ruledtabular}
\end{table*}
Figure~\ref{fig:appendix_team_radial} shows the corresponding team-separated radial curves. The ball-centered drift velocity remains present when the two sides are treated separately. Although the two teams differ in overall drift velocity and in the detailed form of $D(r)$, the main features of the full-game curves are retained.

The two teams in the same game do not always have the same mean inward drift velocity: the difference is largest in Game~1 ($26\%$) and
Game~2 ($17\%$), whereas it is below $9\%$ in each of the remaining
games. The detailed radial dependence can also differ between the two
sides, as reflected by the wider range of $D(r)$ correlations, from
$0.29$ to $0.97$. In contrast, the team-resolved $K(x,y)$ maps
retain similar broad spatial structures in every game, with cell-by-cell
Pearson correlations ranging from $0.904$ to $0.988$. Thus the
large-scale spatial field is not an artifact of averaging two unrelated
team-specific fields, although both the overall drift velocity and the
detailed radial dependence can contain substantial team- and
game-dependent variation.

\section{Player-density test}
\label{app:player_density}

\begin{figure}[t]
\centering
\includegraphics[width=0.95\linewidth]{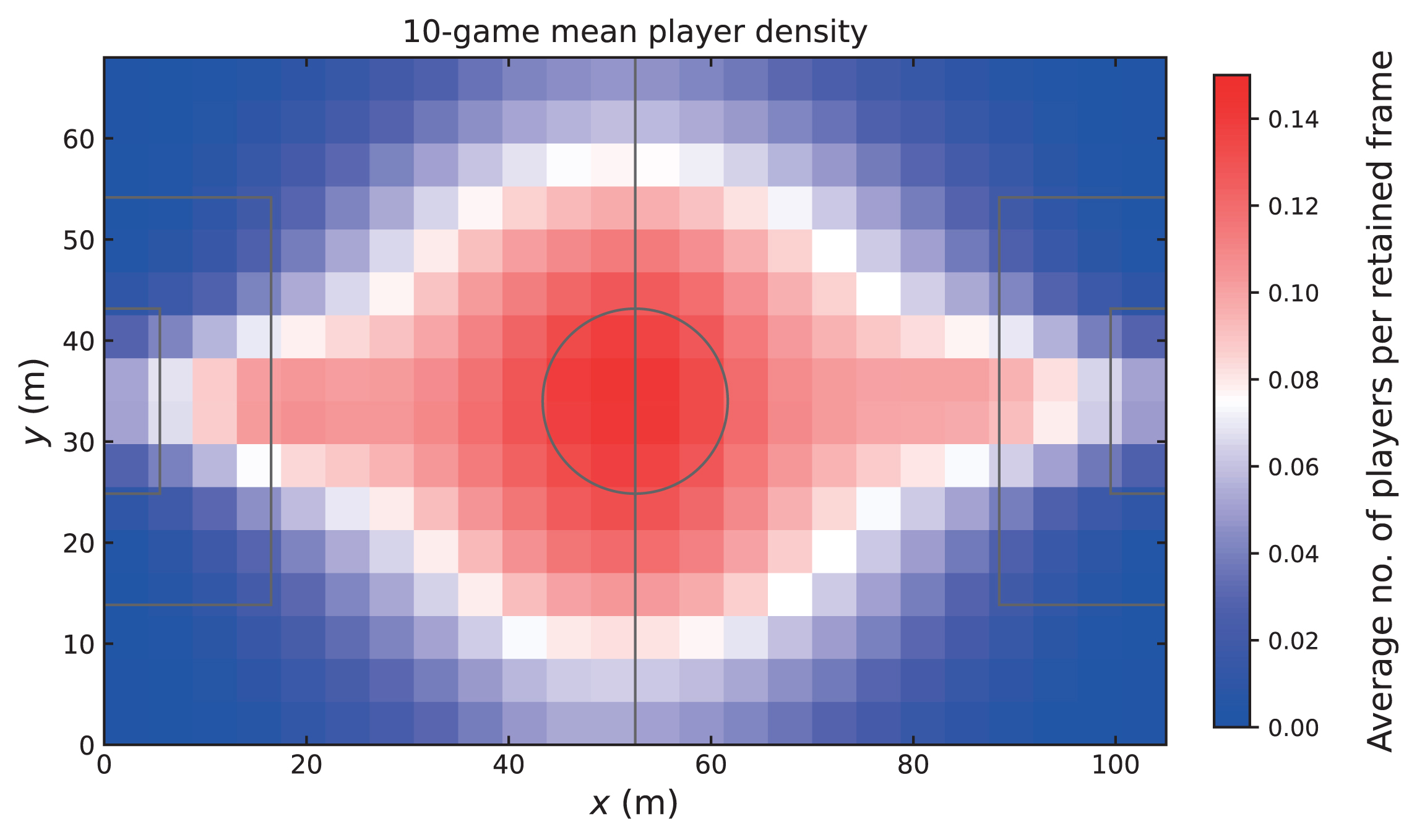}
\caption{Ten-game player-density map, including the goalkeepers and the
player nearest to the ball, on the same $24\times16$ pitch grid as
$K(x,y)$. The same count-weighted Gaussian smoothing as in the other
maps was used.}
\label{fig:appendix_player_density}
\end{figure}

One can also ask whether the central depression and goalmouth suppression
in $K(x,y)$ could be related simply to the spatial distribution of the
players. If this were the case, a similar feature should appear in a
position-count map. We constructed the player-density map shown
in Fig.~\ref{fig:appendix_player_density} from the same retained open-play
frames as the drift velocity analysis, but counting all players present in each
frame. Both goalkeepers and the player nearest to the ball are
included. The map records where players are located on the pitch, rather
than how they move toward the ball.

The largest density occurs at the center of the pitch, to within the grid resolution. Thus the central depression in $K(x,y)$ is not caused by a lack of players in the central region. Players are often concentrated there, but their average motion has a smaller component toward the ball. Likewise, near the goalmouths, the negative values of $K(x,y)$ depend on the direction of player motion and do not follow from the density map. This is because the player-density map and $K(x,y)$ are different quantities. The density map counts player positions on the pitch, whereas $K(x,y)$ averages the component of player velocity toward the ball when the ball lies in a given pitch cell.

\section{Goalkeeper effects and half-match variation}
\label{app:goalkeeper_half}

The calculations reported in the main text exclude the two goalkeepers in
each game. This choice reflects their highly specialized and positionally
constrained role, which differs qualitatively from that of the outfield
players.
To test the influence of this choice, we repeated the calculation after
restoring the two goalkeepers in each game while leaving all other analysis
choices unchanged, including exclusion of the nearest player and the
$r\ge2$~m cutoff.

\begin{table*}[t]
\caption{Goalkeeper-inclusion test for the ten games. The two
goalkeepers, normally excluded, are restored while all other analysis
choices are unchanged. The percentage change is the decrease in the global
mean drift velocity when the goalkeepers are included. The last column
gives the Pearson correlation coefficient between the goalkeeper-excluded
and goalkeeper-included symmetry-averaged $K(x,y)$ maps.}
\label{tab:goalkeeper_check}
\begin{ruledtabular}
\begin{tabular}{lcccc}
Game & \begin{tabular}[c]{@{}c@{}}Goalkeepers excluded\\ Mean $D_i(t)$ ($\mathrm{m\,s^{-1}}$)\end{tabular} & \begin{tabular}[c]{@{}c@{}}Goalkeepers included\\ Mean $D_i(t)$ ($\mathrm{m\,s^{-1}}$)\end{tabular} & Decrease & $K(x,y)$ map corr. \\
\hline
Game 1  & $0.707$ & $0.642$ & $9.2\%$ & $0.9995$ \\
Game 2  & $0.756$ & $0.689$ & $8.9\%$ & $0.9996$ \\
Game 3  & $0.823$ & $0.748$ & $9.2\%$ & $0.9998$ \\
Game 4  & $1.095$ & $1.007$ & $8.0\%$ & $0.9944$ \\
Game 5  & $1.104$ & $1.016$ & $8.0\%$ & $0.9968$ \\
Game 6  & $0.917$ & $0.837$ & $8.7\%$ & $0.9973$ \\
Game 7  & $1.020$ & $0.940$ & $7.8\%$ & $0.9948$ \\
Game 8  & $0.921$ & $0.846$ & $8.1\%$ & $0.9969$ \\
Game 9  & $0.932$ & $0.860$ & $7.8\%$ & $0.9989$ \\
Game 10 & $0.609$ & $0.560$ & $8.1\%$ & $0.9989$ \\
\end{tabular}
\end{ruledtabular}
\end{table*}

Table~\ref{tab:goalkeeper_check} compares the goalkeeper-excluded
calculation used in the main text with a calculation in which the two
goalkeepers are restored. It reports the change in the global mean $D_i(t)$ and the
correlation between the corresponding symmetry-averaged $K(x,y)$ maps.
The global mean values are also compared graphically in
Fig.~\ref{fig:appendix_goalkeeper_half}(a).

Restoring the goalkeepers decreases the global mean inward drift velocity
by between $7.8\%$ and $9.2\%$ across the ten games, with an average
decrease of about $8.4\%$. This is consistent with the positionally
constrained motion of goalkeepers. The spatial pattern changes much less:
the correlations between the goalkeeper-excluded and
goalkeeper-included $K(x,y)$ maps range from $0.9944$ to $0.9998$.
Thus restoring the goalkeepers reduces the mean inward drift velocity but has
little effect on the overall spatial form of $K(x,y)$.

For completeness, we also examine the goalkeepers as a separate playing
role using the same Games~3--10 used in the role-resolved analysis.
This comparison is distinct from the preceding ten-game
goalkeeper-inclusion test: here the goalkeeper-only results and the
outfield-player reference are both equal-weight averages over the same
eight games.

\begin{figure*}[t]
\centering
\includegraphics[width=0.9\linewidth]{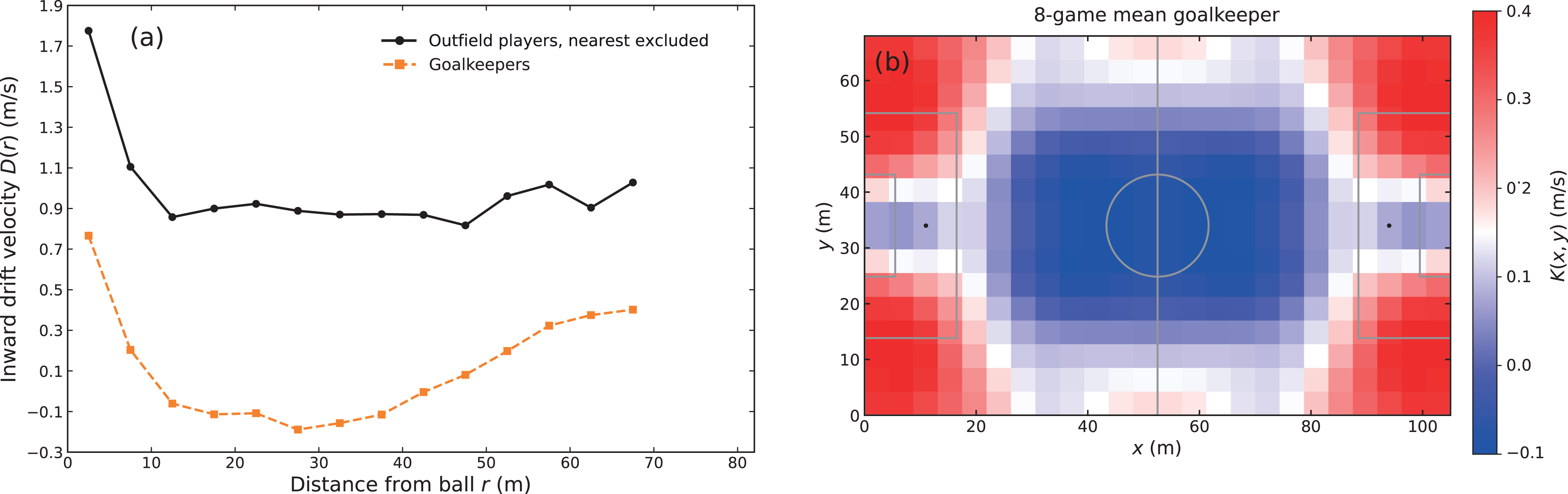}
\caption{Goalkeeper-specific drift velocity for the eight
games used for the role-resolved analysis.
(a) Equal-weight mean radial drift velocity $D(r)$ for the
outfield players and for the goalkeepers alone. Solid black curve: the
two goalkeepers are excluded and, among the remaining outfield players,
the player nearest to the ball at each frame is also excluded. Orange dashed 
curve: only the two goalkeepers in each game are included. Both curves are
averaged over the same eight games and use the same open-play filtering,
$r\ge2$~m cutoff, and radial binning.
(b) Equal-weight mean goalkeeper-only $K(x,y)$ field after the same symmetry averaging and count-weighted
Gaussian smoothing used for the other $K(x,y)$ maps. Positive values
denote mean motion toward the ball and negative values mean motion away
from it.}
\label{fig:goalkeeper_only}
\end{figure*}

\begin{figure}[t]
\centering
\includegraphics[width=0.7\linewidth]{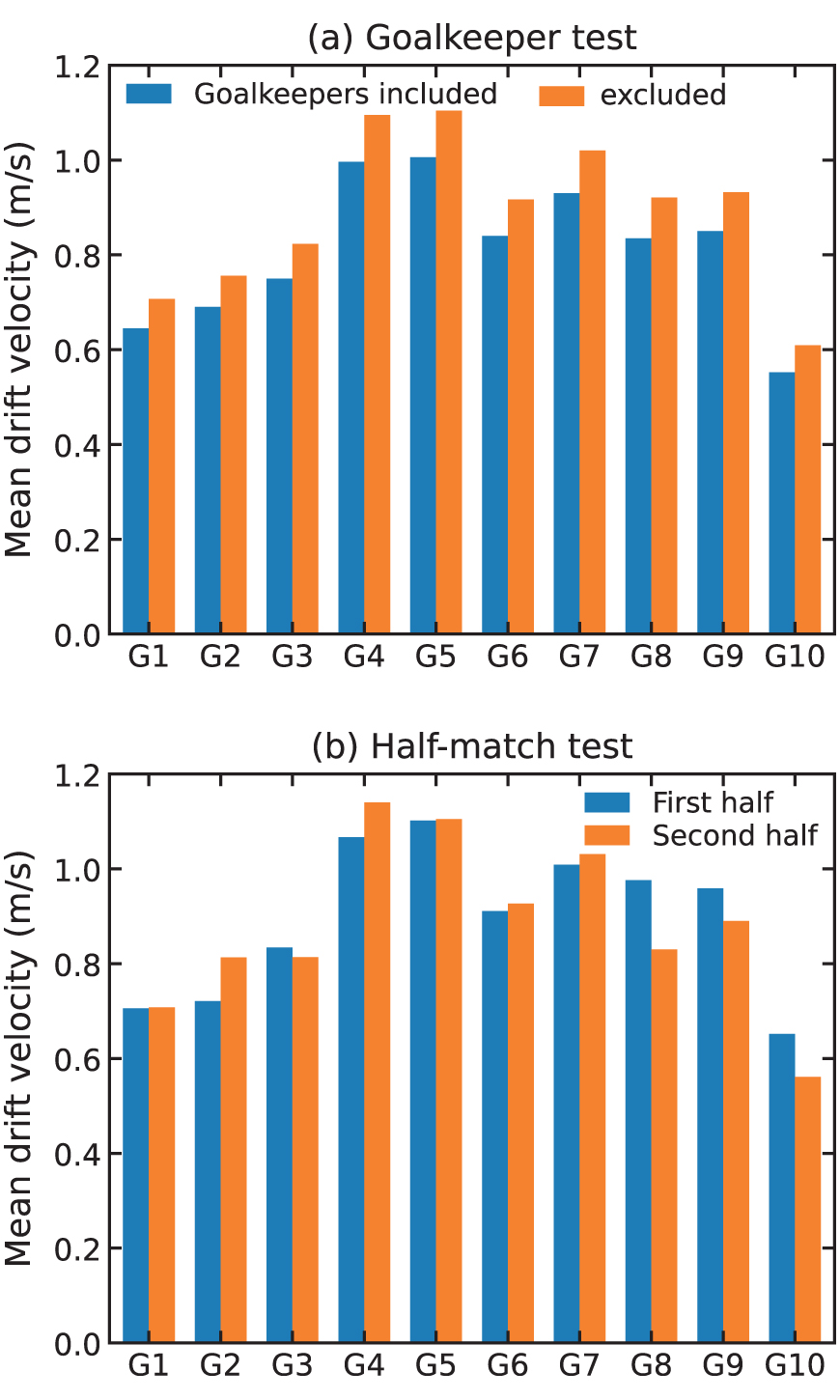}
\caption{Goalkeeper-inclusion and half-match tests for the mean inward
drift velocity. (a) Global mean inward drift velocity with the goalkeepers excluded and
with the two goalkeepers restored. (b) Global mean inward drift velocity
calculated separately for the first and second halves of each game.}
\label{fig:appendix_goalkeeper_half}
\end{figure}

Figure~\ref{fig:goalkeeper_only}(a) compares the eight-game mean $D(r)$
for the outfield-player sample with that for the goalkeepers alone. The outfield-player curve is restricted to Games~3--10 and excludes both
goalkeepers and the player nearest to the ball. The goalkeeper-only curve lies below the
outfield-player curve over most of the distance range and becomes
negative at intermediate player--ball separations, approximately
$15$--$40~\mathrm{m}$. At larger separations it becomes positive again.
This behavior is consistent with the reduction in the global mean drift velocity
seen when the goalkeepers are restored in the ten-game test above.

Panel~(b) shows the corresponding eight-game mean goalkeeper-only
$K(x,y)$ field. When the ball is in a broad central region of the pitch,
the mean goalkeeper drift velocity is negative, whereas it is positive when the
ball is closer to either end. This pattern is consistent with the
specialized positioning of goalkeepers: when the ball lies centrally,
their motion near their own goal has, on average, a component away from
the instantaneous ball position. When the ball is closer to either end,
the mean goalkeeper inward drift velocity becomes positive. The
goalkeeper-only field therefore differs clearly from that of the
outfield players. 

\begin{table*}[t]
\caption{Half-match comparison of the global mean inward drift velocity.
The listed times are the open-play times remaining after restart windows
and other constrained phases of play are removed. The mean values use the goalkeeper exclusion, nearest-player exclusion,
and $r\ge2$~m cutoff used in the main text.
The percentage difference is
$100(D_2-D_1)/[(D_1+D_2)/2]$, where $D_1$ and $D_2$ are the
first- and second-half means, respectively.}
\label{tab:half_check}
\begin{ruledtabular}
\begin{tabular}{lcccccl}
Game & \begin{tabular}[c]{@{}c@{}}First-half time\\ ($\mathrm{min}$)\end{tabular} & \begin{tabular}[c]{@{}c@{}}First half\\ mean $D_i(t)$\\ ($\mathrm{m\,s^{-1}}$)\end{tabular} & \begin{tabular}[c]{@{}c@{}}Second-half time\\ ($\mathrm{min}$)\end{tabular} & \begin{tabular}[c]{@{}c@{}}Second half\\ mean $D_i(t)$\\ ($\mathrm{m\,s^{-1}}$)\end{tabular} & Difference & Comment \\
\hline
Game 1  & $17.52$ & $0.706$ & $22.43$ & $0.708$ & $+0.3\%$  & nearly unchanged \\
Game 2  & $23.28$ & $0.721$ & $14.47$ & $0.813$ & $+12.1\%$ & stronger second half \\
Game 3  & $16.19$ & $0.834$ & $19.36$ & $0.814$ & $-2.4\%$  & slightly weaker second half \\
Game 4  & $23.60$ & $1.067$ & $14.72$ & $1.140$ & $+6.6\%$  & stronger second half \\
Game 5  & $15.59$ & $1.102$ & $14.73$ & $1.105$ & $+0.3\%$  & nearly unchanged \\
Game 6  & $17.15$ & $0.911$ & $9.57$  & $0.927$ & $+1.7\%$  & slightly stronger second half \\
Game 7  & $16.14$ & $1.009$ & $15.88$ & $1.031$ & $+2.2\%$  & slightly stronger second half \\
Game 8  & $22.96$ & $0.976$ & $13.89$ & $0.830$ & $-16.2\%$ & weaker second half \\
Game 9  & $24.41$ & $0.959$ & $16.20$ & $0.890$ & $-7.5\%$  & weaker second half \\
Game 10 & $21.86$ & $0.652$ & $19.70$ & $0.561$ & $-15.0\%$ & weaker second half \\
\end{tabular}
\end{ruledtabular}
\end{table*}

The half-match results are shown graphically in
Fig.~\ref{fig:appendix_goalkeeper_half}(b), using the values listed in
Table~\ref{tab:half_check}. The half-to-half differences range from
$-16.2\%$ to $+12.1\%$, with a mean absolute difference of about
$6.4\%$. There is no systematic tendency for the second half to exhibit
either stronger or weaker inward drift velocity: some games show an increase,
others a decrease, and several change only slightly. Most importantly,
the mean inward drift velocity remains positive in both halves of all ten games.
Thus the persistent inward drift velocity does not depend on averaging the two
halves of a match together, although its magnitude naturally varies
between halves.

\begin{figure}[t]
\centering
\includegraphics[width=\linewidth]{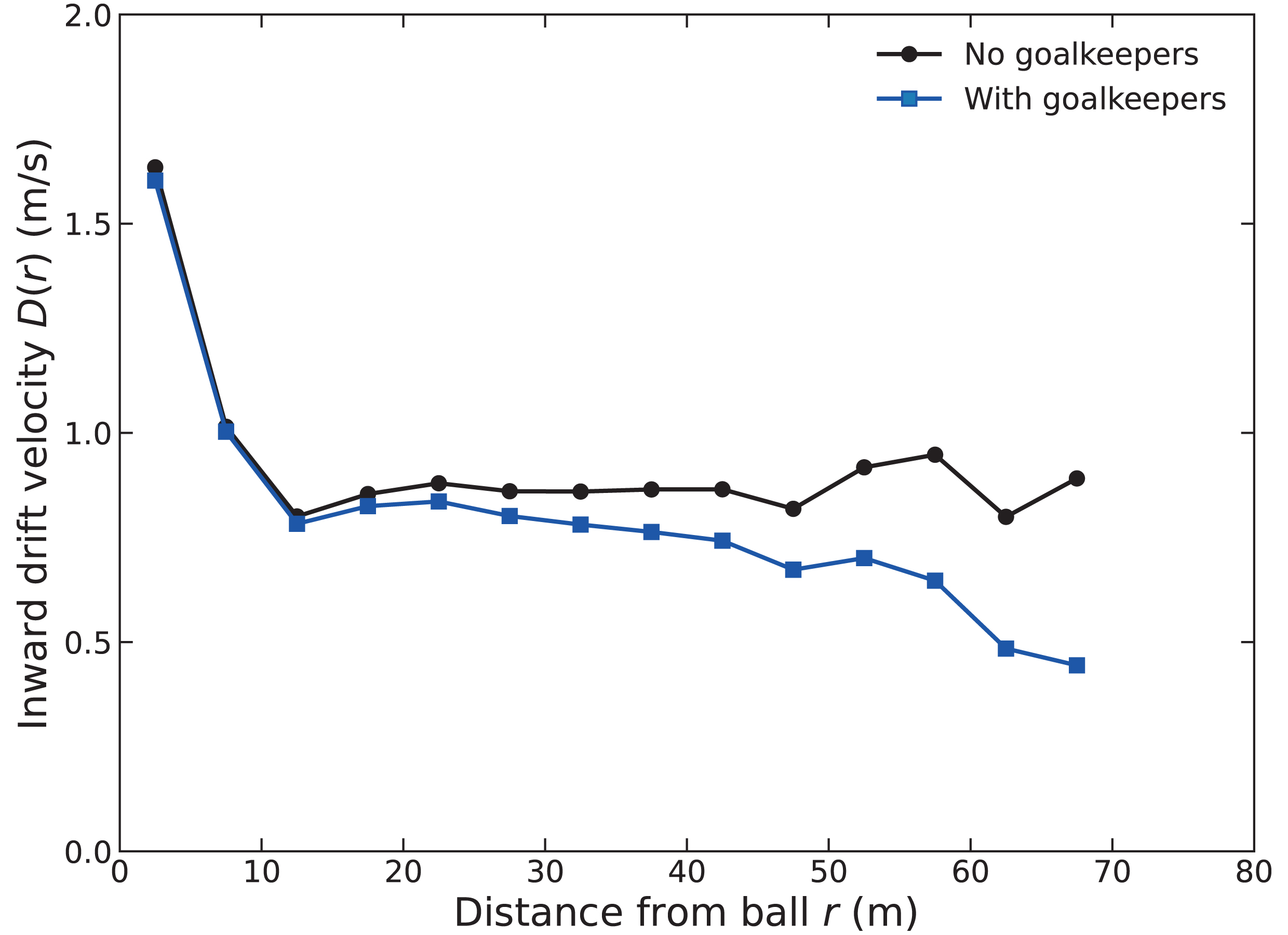}
\caption{Ten-game mean radial drift velocity $D(r)$ with the two
goalkeepers excluded (solid black circles) and included (solid blue squares).
In both calculations, the player nearest to the ball is excluded. The
same open-play filtering, $r\ge2~\mathrm{m}$ cutoff, and $5~\mathrm{m}$
distance intervals are used. Positive values denote mean motion toward
the ball.}
\label{fig:gk_Dr_ten_game}
\end{figure}

Finally, Fig.~\ref{fig:gk_Dr_ten_game} compares the ten-game mean
$D(r)$ with the goalkeepers excluded and included. In both calculations,
the player nearest to the ball is excluded. Restoring the two goalkeepers
has relatively little effect at short and intermediate distances but
progressively reduces the mean inward drift velocity at large $r$,
producing a stronger decrease beyond about $40$--$50~\mathrm{m}$. This stronger large-distance decrease is consistent with
Fig.~\ref{fig:goalkeeper_only}(a), where the goalkeeper-only $D(r)$
lies well below the corresponding outfield-player curve over much of the
distance range.

We also recalculated $K(x,y)$, averaged over all ten games, with the
goalkeepers included and found no appreciable change in its overall
spatial form.

\FloatBarrier

\bibliographystyle{apsrev4-2}
\bibliography{soccer_gravity_refs}

\end{document}